\documentclass{aa}
\usepackage{graphicx}
\usepackage{txfonts}

\begin{document}

\title{A new method to perform data/model comparison in {\it Fermi}-LAT analysis}

\author{P. Bruel}

\institute{Laboratoire Leprince-Ringuet, CNRS, Ecole polytechnique, Institut Polytechnique de Paris, F-91128 Palaiseau, France\\
  \email{Philippe.Bruel@llr.in2p3.fr}
}

\date{Received 15 June 2021; accepted 15 September 2021}

\abstract
{The analysis of {\it Fermi} Large Area Telescope (LAT) gamma-ray data in a given Region Of Interest (RoI) usually consists of performing a binned log-likelihood fit in order to determine the sky model that, after convolution with the instrument response, best accounts for the distribution of observed counts.}
{While tools are available to perform such a fit, it is not easy to check the goodness-of-fit. The difficulty of the assessment of the data/model agreement is twofold. First of all, the observed and predicted counts are binned in three dimensions (two spatial dimensions and one energy dimension) and comparing two 3D maps is not straightforward. Secondly, gamma-ray source spectra generally decrease with energy as the inverse of the energy square. As a consequence the number of counts above several GeV generally falls into the Poisson regime, which precludes performing a simple $\chi^2$ test.}
{We propose a method that overcomes these two obstacles by producing and comparing spatially integrated count spectra for data and model at each pixel of the analysed RoI. The comparison is performed following a log-likelihood approach that extends the $\chi^2$ test to histograms with low statistics. This method can take into account likelihood weights that are used to account for systematic uncertainties.}
{We optimize the new method so that it provides a fast and reliable tool to assess the goodness-of-fit of {\it Fermi}-LAT data and we use it to check the latest gamma-ray source catalog on 10~years of data.}
{}


\maketitle
%

\section{Introduction} \label{sec:intro}

Since its launch in June 2008, the {\it Fermi} Large Area Telescope~\citep[LAT,][]{latinstrument} has been continuously observing the gamma-ray sky in the energy range between 30~MeV and 2~TeV. The {\it Fermi}-LAT data are public\footnote{\url{https://fermi.gsfc.nasa.gov/ssc}} and have been widely used by the gamma-ray community. The {\it Fermi}-LAT latest general catalog, 4FGL-DR2~\citep{4FGL,4FGL-DR2}, reports almost 5800 sources, among which are hundreds of pulsars, tens of supernovae remnants and pulsar wind nebulae, and thousands of blazars.

The usual way to analyse LAT data in a given Region of Interest (RoI) is to bin the data in a three dimensional map (two spatial dimensions and one energy dimension~\footnote{A typical example is a $12\degr\times 12\degr$ map with a pixel size of $0.1\degr$ and 10 logarithmically spaced bins per decade in energy.}) and then search for the sky model that best predicts the number of gamma rays observed in this RoI. The sky model is a list of gamma-ray sources, whose position, spatial nature (point-like or extended), as well as their energy spectrum are provided. The {\it Fermi}-LAT analysis package, {\tt Fermitools}\footnote{\url{https://fermi.gsfc.nasa.gov/ssc/data/analysis/documentation}}, is used to convolve the emission of each source with the Instrument Response Functions~\citep[IRFs,][]{IRFs} in order to predict the number of observed counts. As a consequence, one of the main steps of the analysis is generally a log-likelihood fit that finds the source spectral parameters that give the best agreement between the observed and predicted 3D count maps.

After the fit has converged, it is important to check that the obtained agreement is satisfactory but the user faces the difficulty of comparing two 3D maps. The simplest way would be to define an energy band and produce the corresponding 2D residual ($\mathrm{data}-\mathrm{model}$) count map. But the energy band definition and the analysis of the residual map is not straightforward because of the energy dependence of the Point Spread Function (PSF) of the instrument: its 68\% containment angle varies from $5\degr$ at 100~MeV to $0.1\degr$ at 30~GeV~\citep{latinstrument,pass8}~\footnote{\url{https://www.slac.stanford.edu/exp/glast/groups/canda/lat_Performance.htm}}. As a consequence the binning of the residual map should depend on the energy range definition. Furthermore, since the spectral characteristics of a potential data/model mismatch are unknown, it is not obvious to choose {\it a priori} an energy range, leading often to look at several energy bands.

The {\tt Fermitools} provide a method to quantify the data/model agreement which is frequently used. It consists of computing a Test Statistic (TS) map: at each pixel, the presence of an additional source is tested by computing twice the difference in log-likelihood obtained with and without the source. A $\mathrm{TS}=25$ corresponds to $\sim 4\sigma$ significance~\citep{mattox}. The drawbacks of this method are twofold: it is computationally intensive and, above all, it is not sensitive to negative deviations (data<model) because the flux of the additional source is bound to be positive. Another public analysis package, {\it Fermipy}~\citep{fermipy}, offers a simplified version of the TS maps that is much faster but is still blind to negative deviations.

In this paper we propose a method to perform data/model comparison that overcomes the mentioned problems. As in the case of residual or TS maps, we want to provide spatial information. So the goal is to quantify the level of deviation between data and model at each pixel of the RoI. Since we go from 3D count maps to a 2D deviation map, it implies that the spectral information must be fully utilised in the process of the deviation assessment. So, for each pixel of the RoI, the method consists of building and comparing data and model count spectra. We present in Section~\ref{sec:integratedcountspectra} a simple way to build spatially-integrated count spectra that takes into account the PSF of the instrument. Since it is not always possible to perform a simple $\chi^2$ test to compare these count spectra, we apply a log-likelihood approach that is presented in Section~\ref{sec:deviationprobability} and its extension to take into account systematic uncertainties is presented in Section~\ref{sec:wLL}. We optimize and verify this new method in Section~\ref{sec:PScalibration} and the results obtained when using it to check the 4FGL-DR2 catalog are given in Section~\ref{sec:catXcheck}.

\section{PSF-like integrated count spectra} \label{sec:integratedcountspectra}

Since we want to be as sensitive as possible to discrepancies between data and model, it is useful to list the possible causes for such discrepancies in order to find the optimal way to create the count spectra that we will compare in Section~\ref{sec:deviationprobability}. The two possibilities are that one source is missing in the model or that it is mismodelled. In both cases, the resulting discrepancy depends on the point-like or extended nature of the source. In the point-like case, the region of the discrepancy corresponds to the PSF and, as a consequence, its size follows the energy dependence of the PSF.

The LAT PSF 68\% containment angle can be parameterized as $p_0 (E/100~\mathrm{MeV})^{-p_1} \oplus p_2$, the addition being in quadrature~\citep{PSFDETERMINATION}. For LAT PASS~8 {\tt SOURCE} class events, $p_0$ and $p_2$ are about 5 and $0.1\degr$ and $p_1 = 0.8$.

If the source is slightly extended ({\it e.g.} $0.5\degr$), the size of the discrepancy region can be modelled by the same parameterization, replacing $p_2$ by the sum of $0.1\degr$ and the extension of the source. For diffuse sources like the Galactic diffuse emission, one can use an even larger $p_2$.

\begin{figure}[ht]
  \centering
  \includegraphics[width=9.5cm]{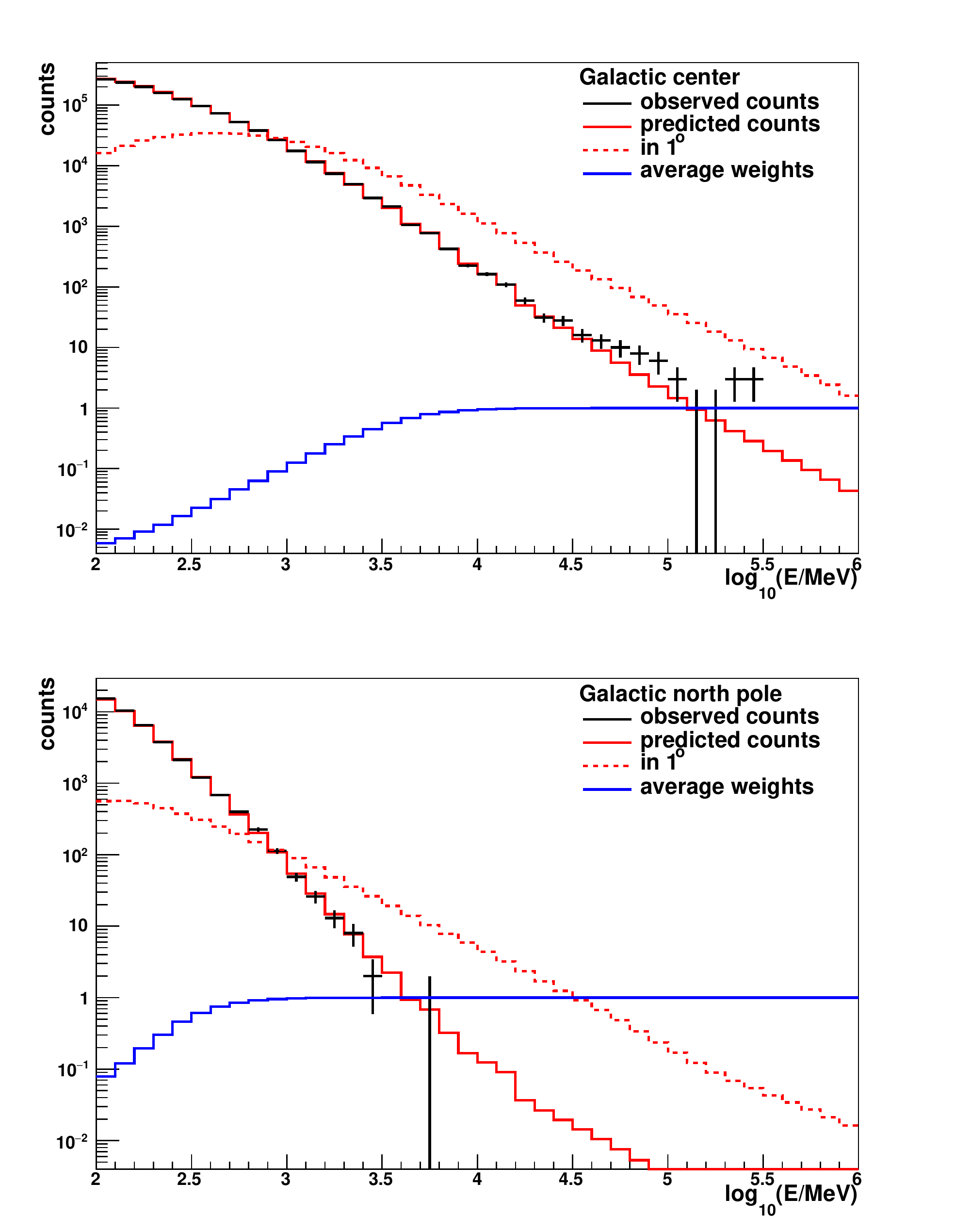}
  \caption{PSF~68\% integrated count spectra at the center pixel of the RoIs centered on the Galactic center (top) and the galactic North pole (bottom) for data (black) and model (red). The dashed red line shows for comparison the model count spectra integrated over $1\degr$. The average likelihood weights, introduced in Section~\ref{sec:wLL}, are also shown.}
  \label{fig:integratedcountspectra}
\end{figure}

For these reasons, we propose to use an energy dependent distance of the form $p_0 (E/100~\mathrm{MeV})^{-p_1} \oplus p_2$ to build for each pixel of the RoI the spatially integrated count spectra for both data and model. We start with the {\tt SOURCE} class PSF~68\% containment angle but the definition of the integration region will be revisited and optimized in Section~\ref{sec:PScalibration}. Examples of PSF~68\% integrated count spectra are shown in Figure~\ref{fig:integratedcountspectra}. For simplicity's sake, the spatial integration is performed directly on the data and model 3D count maps used in the fit (by summing over the pixels within the energy dependent distance). This choice allows us to compare data and model using the closest information to the one used in the fit.

Regarding the energy binning of the count spectra, it is useful to consider the expected spectral features of potential data/model deviations. The known gamma-ray sources do not exhibit any narrow peak or dip in their spectra in the {\it Fermi}-LAT energy range and we thus do not expect particularly sharp data/model deviations from either a spectrally mismodelled source or a missing source. Any deviation would be smeared by the energy resolution of the LAT, which is 10\% at 1~GeV, corresponding to a 68\% containment interval of about 0.1 in $\log_{10}{E}$. As a result, it is not useful to use a bin smaller than 0.1. On the contrary, a larger binning may increase the deviation sensitivity by increasing the statistics in each bin. The choice of the optimal bin size will be investigated in Section~\ref{sec:PScalibration}.

\section{Estimation of the deviation probability} \label{sec:deviationprobability}

In order to assess the level of deviation between the data and model integrated count spectra, we want to compute the p-value, that is to say the probability that the statistical fluctuations can reach a level of deviation as large as the one observed in the data, under the assumption that the model represents the data.

If all the bins of the count spectra were in the Gaussian regime, we could simply perform a $\chi^2$-test to estimate the p-value. But because of the general $E^{-\gamma}$ power-law spectrum of gamma-ray sources (with photon indices $\gamma$ between 1 and 3 for most of the sources as well as for the background) and the energy dependence of the PSF~68\% selection, the integrated count spectra fall steeply with energy and the numbers of counts at high energy are generally in the Poisson regime, which precludes performing a $\chi^2$-test.

In order to overcome this limitation, we adopt a log-likelihood approach and define the following random variable:

\begin{equation} \label{eq:loglike}
L = -\sum_k \log \mathcal{P}(x_k,m_k)
\end{equation}
where $\mathcal{P}$ is the Poisson probability and $x_k$ are independent random Poisson variables of mean $m_k$, the spatially integrated number of model counts in the spectral bin $k$. The p-value is the integral of the probability distribution function (pdf) of $L$ above $L_\mathrm{data} = -\sum_k \log \mathcal{P}(n_k,m_k)$, the value obtained with the data integrated count spectra $n_k$. In other words, the p-value is the $L$ complementary cumulative distribution function (CCDF) at $L_\mathrm{data}$.

It is common to compute expected likelihood distributions by performing simulations but it can be very computationally intensive, especially when requiring the resulting statistical fluctuations to be below the percent level. Here we choose to compute the $L$ pdf in an iterative way, as fully described in~Appendix~\ref{app:iterativecomputation}. When the number of counts in part of the spectrum is large enough, it is possible to optimize the computation by taking advantage of the $\chi^2$ approximation of the log-likelihood. For the spectral bins with Gaussian statistics, the Poisson probability of parameter $m_k$ can be replaced with a Gaussian with mean and variance equal to $m_k$. Ignoring the constant term, we have:

\begin{equation} \label{eq:chi2}
L_\mathrm{Gaus} = \frac{1}{2} \sum_k \frac{(x_k-m_k)^2}{m_k}
\end{equation}
As a consequence, the $L_\mathrm{Gaus}$ pdf can be simply derived from the $\chi^2$ distribution with a number of degrees of freedom equal to the number of spectral bins with Gaussian statistics.

In order to compute the $L$ pdf, we thus start by sorting the counts in decreasing order, then compute the $L_\mathrm{Gaus}$ pdf corresponding to all the bins with a number of counts greater than 100 and finally perform the iterative computation over the remaining bins. We note that this p-value computation provides a simple extension of the $\chi^2$-test to histograms with counts in the Poisson regime.

We define PS~\footnote{The name PS was chosen because P can stand for both p-value and PSF and also because the output map name, PS map, sounds close to TS map.}, the data/model deviation estimator, as:

\begin{equation} \label{eq:absPS}
  \mathrm{|PS|} = -\log_{10}(\text{p-value})
\end{equation}
and give it the sign of the sum of the residuals in sigma units:

\begin{equation} \label{eq:signPS}
  \mathrm{sign(PS)} = \mathrm{sign} \left( \sum_k (n_k-m_k)/\mathrm{max} \left(1,\sqrt{m_k} \right) \right)
\end{equation}
which allows us to estimate whether the deviation is positive (data>model) or negative (data<model). To our knowledge, this is the first time that such an estimator is proposed.

The advantage of using the logarithm of the p-value rather than converting it into $\sigma$ units is that, when considering the maximum value of a PS map, the correction for the number of trials is simply done by subtracting the logarithm of the number of pixels in the map. The $3,4$ and $5\sigma$ thresholds correspond to $\mathrm{PS}=2.57,4.20$ and $6.24$, respectively. For a typical $100 \times 100$ pixels map, assuming that the PS are independent, the $3,4$ and $5\sigma$ thresholds correspond to uncorrected $\mathrm{PS}=6.57,8.20$ and $10.24$, respectively.

\section{Systematic uncertainty handling with weighted log-likelihood } \label{sec:wLL}

Log-likelihood weights have been introduced in the {\it Fermi}-LAT general catalog analysis in order to account for systematic uncertainties, especially those coming from the modelling of the diffuse emission~\citep{4FGL}. This is done by performing the spectral fit with the following definition of the log-likelihood~\citep{loglikeweights}:

\begin{equation} \label{eq:catwloglike}
\log \mathcal{L} = \sum_{i,j,k} W_{ijk} (N_{ijk} \log M_{ijk} -M_{ijk})
\end{equation}
where the indices $i,j,k$ run over the 3D maps and $W,N$ and $M$ are the map pixel weight, number of observed and predicted counts, respectively. The {\it Fermi}-LAT catalog analysis uses data-based weights corresponding to a level of systematic uncertainty of 3\% \citep[see Appendix-B of][for more details]{4FGL}.

We note that in the Gaussian regime, the Poisson contributions in Equation~\ref{eq:catwloglike} can be replaced with Gaussian probabilities:

\begin{equation} \label{eq:wloglike_gaus}
\frac{1}{2}W_{ijk}(N_{ijk}-M_{ijk})^2/M_{ijk} = \frac{1}{2}(N_{ijk}-M_{ijk})^2/(M_{ijk}/W_{ijk})
\end{equation}
which highlights that the effect of the weights is to increase the variance by $1/W_{ijk}$.

If log-likelihood weights are used in the spectral fit, they also have to be taken into account when assessing the goodness-of-fit. The first thing to do is thus to also introduce weights in the definition of the random variable $L$ used to compute the p-value and Equation~\ref{eq:loglike} becomes:

\begin{equation} \label{eq:wloglike}
L = -\sum_k w_k\log \mathcal{P}(x_k,m_k)
\end{equation}
Since $m_k$ is the sum of predicted counts in the integration region, its variance is the sum of the pixel variances. Following Equation~\ref{eq:wloglike_gaus}, we interpret $m_k/w_k$ as the expected variance of $m_k$ and thus define $w_k$ such that $m_k/w_k = \sum_{ij} M_{ijk}/W_{ijk}$, where the sum runs over the integration region corresponding to the spectral bin $k$. Since we adopt the data-based weights as in the 4FGL-DR2 analysis, the average is actually weighted with the data counts rather than the model counts: $n_k/w_k = \sum_{ij} N_{ijk}/W_{ijk}$. When $n_k=0$, $w_k$ is set to 1. Examples of such average weights are shown in Figure~\ref{fig:integratedcountspectra}.

Although Equation~\ref{eq:wloglike} accounts for the relative importance of the systematic uncertainties between spectral bins, it does not take into account their absolute meaning. This is simply because the variances of the individual Poisson distributions are unchanged. A clear symptom of this problem is that the p-value is invariant with a global rescaling of the weights.
 
This serious limitation can be naturally overcome for the bins of the count spectrum that follow Gaussian statistics, for which the introduction of the weights modifies Equation~\ref{eq:chi2} as:

\begin{equation} \label{eq:wchi2}
L_\mathrm{Gaus} = \frac{1}{2} \sum_k \frac{(x_k-m_k)^2}{m_k/w_k}
\end{equation}
As a consequence, using the $\chi^2$ approximation $L_\mathrm{Gaus}$ to compute the $L$ pdf in the weighted log-likelihood case allows us to ensure that the absolute meaning of the systematic uncertainty is properly taken into account, as long as the uncertainties of the spectral bins are taken as $\sqrt{m_k/w_k}$ instead of $\sqrt{m_k}$.

In order to study the case of the spectral bins with Poisson statistics, we perform simulations for different levels of systematic uncertainties, whose results are described in Appendix~\ref{app:systwithweights}. The conclusion of this study is that introducing the weights in the definition of $L$ is helpful for large systematic uncertainty and unimportant for small systematic uncertainty, where the needed correction is actually small. In this study considering only the bins with Poisson statistics, the resulting error on PS is within 3\% for systematic uncertainties of the order of 3\%, as currently used in LAT analyses. When computing the PS with all the bins of the count spectra, the true error on PS is actually smaller thanks to the bins in the Gaussian regime, which make up about 50\% of the spectral bins in the analysis of 10~years of LAT data, as reported in the next Section.

Because of the overall positive role of the weights to take into account systematic uncertainty, we decide to keep the weighted version of $L$ to compute the PS. As a consequence, when computing the $L$ pdf, we first compute the $L_\mathrm{Gaus}$ pdf corresponding to all the spectral bins with a number of counts greater than 100, using Equation~\ref{eq:wchi2}, and then perform the iterative computation over the remaining bins using Equation~\ref{eq:wloglike}. Compared to Equation~\ref{eq:signPS}, the PS sign definition is modified as:

\begin{equation} \label{eq:signwPS}
  \mathrm{sign(PS)} = \mathrm{sign} \left( \sum_k (n_k-m_k)/\mathrm{max} \left(1,\sqrt{m_k/w_k} \right) \right)
\end{equation}

\section{PS optimization and calibration} \label{sec:PScalibration}

We use the PS method to assess how well the {\it Fermi}-LAT general catalog describes the whole sky in terms of predicted counts. This verification, named catXcheck, performed on the latest published catalog, {\it i.e.} 4FGL-DR2, gives us the opportunity to measure and optimize the PS sensitivity.

\subsection{catXcheck framework}

It consists of the analysis of 438 RoIs ($120\times120$~pixels, with a pixel size of $0.1\degr$) covering the whole sky. The RoI centers lie on galactic parallels whose latitudes go from $-90$ to 90 with a $10\degr$ step. The longitude step is $10\degr$ for $|b| \leq 30$ and it is 12, 15, 20, 30 and $45\degr$ for $|b|=40$, 50, 60, 70 and 80, respectively.

The analysis of each RoI is performed with the {\tt Fermitools}. We use the same data as in the 4FGL-DR2 catalog analysis, namely Pass~8 SOURCE class data~\citep{pass8,P8R3} collected during the first 10~years of the mission. We select data above 100~MeV whose zenith angle is less than $90\degr$ to avoid Earth's limb contamination. Using the PSF event-type partition improves on average the point source sensitivity but it is not particularly useful when assessing data/model agreement. We thus combine all events in the analysis. We use 10~bins per decade between 100~MeV and 1~TeV. The model comprises:
\begin{itemize}
\item the Galactic diffuse emission and the isotropic template~\footnote{\url{https://fermi.gsfc.nasa.gov/ssc/data/access/lat/BackgroundModels.html}};
\item the Sun and Moon steady emission templates;
\item all point-like and extended sources from 4FGL-DR2, within $5+0.015(\sigma_\mathrm{src}-4)$~degrees of the RoI border, where $\sigma_\mathrm{src}$ is the significance of the source as reported in the catalog.
\end{itemize}
The only free parameters of the spectral fit are the ones of the Galactic diffuse emission (power-law correction) and isotropic (normalization) templates. We use the {\tt P8R3\_SOURCE\_V2} IRFs and energy dispersion is taken into account for the Galactic diffuse emission and the 4FGL sources.

\subsection{PS optimization} \label{sec:PSoptimization}

The PS estimator is designed to detect data/model deviations. A way to quantify and optimize its sensitivity is to create artificial deviations. This is done within the catXcheck framework: for each 4FGL source inside an RoI and more than $1\degr$ away from its border, we remove the source from the model, recompute the total predicted 3D count map and then compute the PS and TS around the position of the source. In order to find the spatial selection parameters as well as the energy binning of the count spectra that maximize the PS, we compare the PS to the TS over the whole set of 4FGL-DR2 sources. We stress that comparing the PS and TS does not imply that we suggest that the PS could replace the TS as a way to quantify the significance of a known gamma-ray source. The only goal of the PS estimator is to search for data/model deviations. As such, it can detect a potential missing source in the source model of an RoI but it is not the optimal way to characterize a known gamma-ray source.

\begin{figure}[ht]
  \centering
  \includegraphics[width=9.5cm]{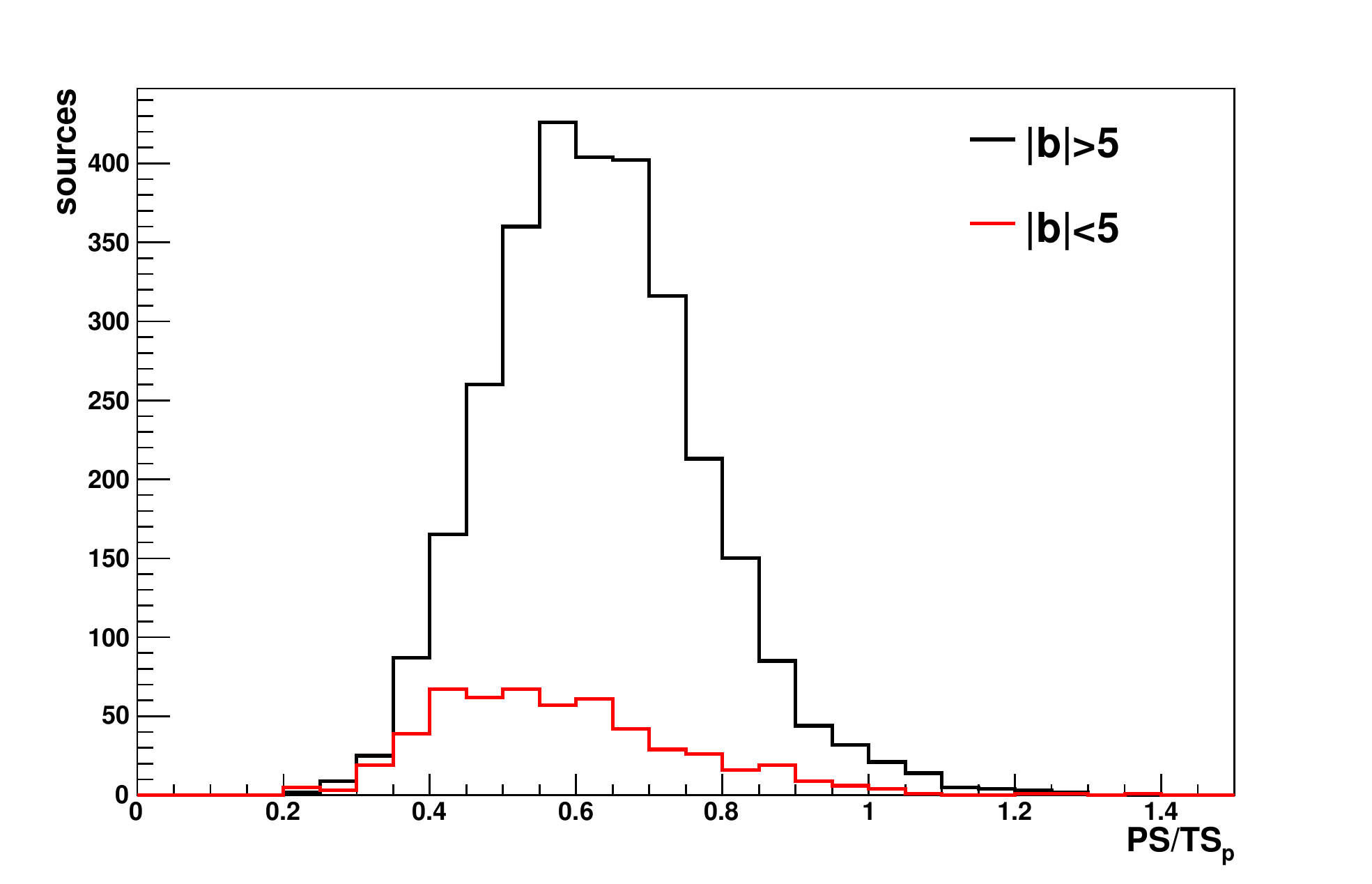}
  \caption{The distribution of the ratio of PS over $\mathrm{TS}_\mathrm{p}$ (TS expressed as $-\log_{10}(\text{p-value})$) for the 4FGL-DR2 sources closer (red) and further than $5\degr$ (black) from the Galactic plane.}
  \label{fig:pstsratio}
\end{figure}

For the integrated count spectra used to compute the PS, we first use a bin width of 0.1 in $\log_{10}{E}$. For the TS computation, the putative source is modelled with a power law with a photon index fixed to 2.3, which is the average photon index of the 4FGL-DR2 sources. So we assume that the TS follows a $\chi^2$ distribution with one degree of freedom. The PS sensitivity can be compared to that of the TS by simply looking at the ratio of the maximum PS to $\mathrm{TS}_\mathrm{p}$, the maximum TS expressed as $-\log_{10}(\text{p-value})$. This ratio (which, for simplicity's sake, we refer to as PS/TS) is shown in Figure~\ref{fig:pstsratio} for sources with $\mathrm{TS}_\mathrm{p}>7$ in ($|b|<5\degr$) and outside ($|b|>5\degr$) the Galactic plane. These distributions are relatively wide with a peak at around 0.6.

The PS is computed on the integrated count spectra and the energy dependent spatial selection is defined with $p_0 (E/100~\mathrm{MeV})^{-p_1} \oplus p_2$. In order to find which parameters maximize the PS sensitivity, we compute the PS/TS ratio on a $p_0,p_1$ grid for $p_2=0.1, 0.15$ and 0.20$\degr$. Above 30~GeV, for a pixel size of $0.1\degr$, these three values of $p_2$ correspond to an integration region of 5, 9 and 13 pixels, respectively.

\begin{figure}[ht]
  \centering
  \includegraphics[width=9.5cm]{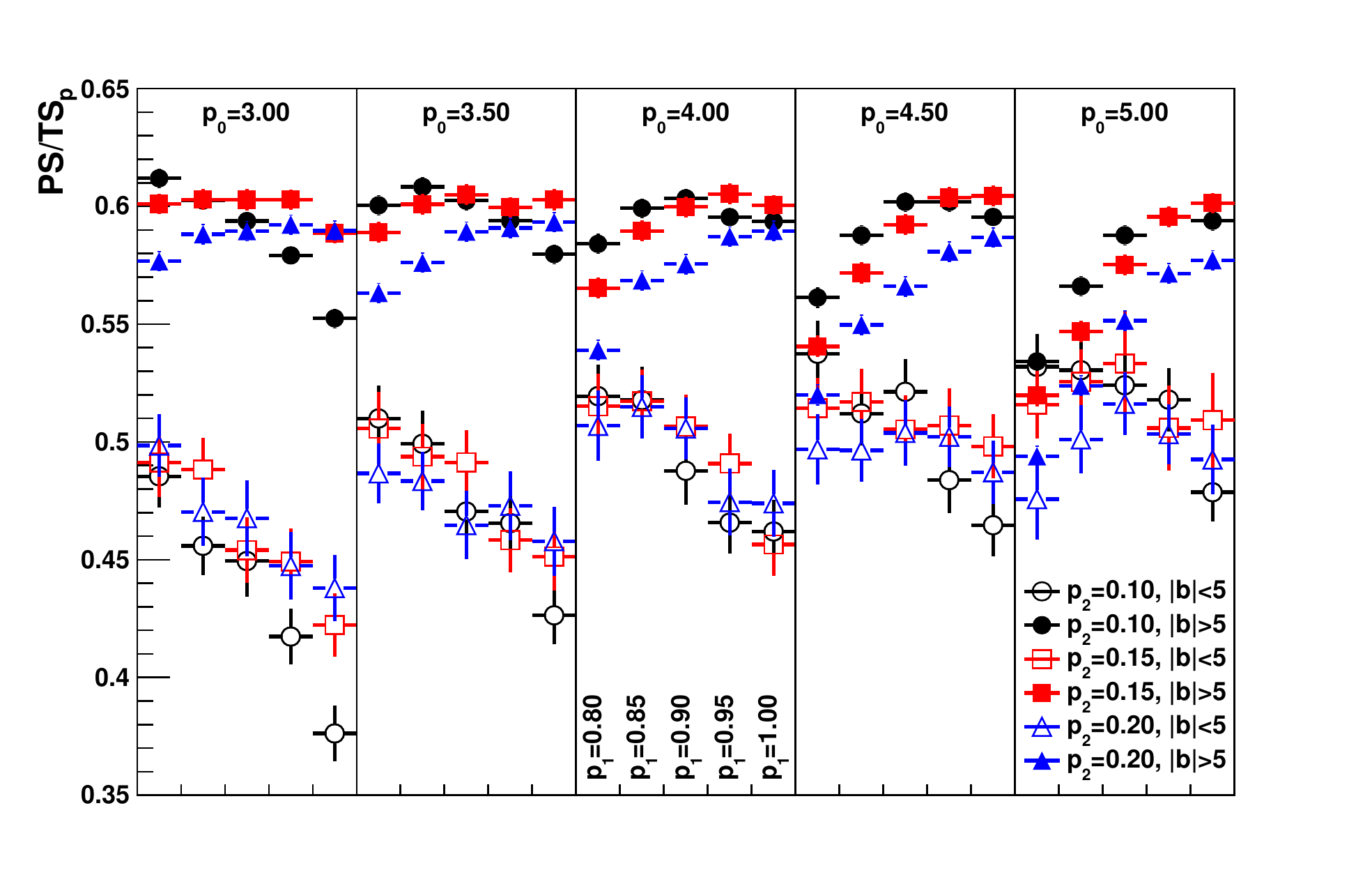}
  \caption{Position of the maximum of the PS/TS distribution of low and high-latitude sources as a function of the spatial selection parameters $p_0$ and $p_1$ for three choices of $p_2=0.10, 0.15$ and $0.20\degr$.}
  \label{fig:psfparoptimization}
\end{figure}

For each configuration, we fit the PS/TS distribution with a log normal in order to estimate the peak position. Figure~\ref{fig:psfparoptimization} shows how this peak position varies with the parameters. There is no significant difference between the results obtained with $p_2=0.1$ and 0.15 but the results are on average worse with $p_2=0.2$. Over most of the $p_0,p_1$ grid, the variation is modest compared to the typical 0.15 width of the distributions.  A precise choice of the parameters is thus not critical and we choose $p_0=4$ and $p_1 = 0.9$. We note the that the dependence of the ratio on the parameters $p_0$ and $p_1$ is opposite in the Galactic plane and away from it: a larger integration region at low energy is preferred in the former case. This is due to the fact that the spectrum of the Galactic diffuse emission is harder in the plane than away from it, whereas the 4FGL-DR2 sources are on average softer. Regarding $p_2$, we choose $0.1\degr$ in order to minimize the level of correlation between pixels, as discussed in the next Section. These parameters, defining the optimized spatial selection, are the ones used to produce the PS/TS ratio distributions of Figure~\ref{fig:pstsratio}.

All these results have been obtained with a TS computed as normally done in the computation of TS maps, that is to say assuming a power-law spectrum. This choice, along with the fact that the PS and TS are computed using the same data and model 3D maps, ensures a fair PS/TS comparison. However $\sim30$\% of the sources in 4FGL-DR2 are better modelled with a curved spectrum so the power-law assumption may bias the optimization of the spatial selection parameters. In order to check this possibility, we perform the same comparison but using the TS reported in the 4FGL-DR2 catalog, which is computed with the curved spectral shape (either a log normal or a subexponentially cutoff power law) when the source spectrum is found to be significantly curved. We find that, compared to the results presented in Figure~\ref{fig:psfparoptimization}, the average value is decreased by $\sim 10$\% but the variation of the PS/TS ratio with the spatial selection parameters is the same, which shows that the optimization procedure was not biased by the power-law assumption.

\begin{figure}[ht]
  \centering
  \includegraphics[width=9.5cm]{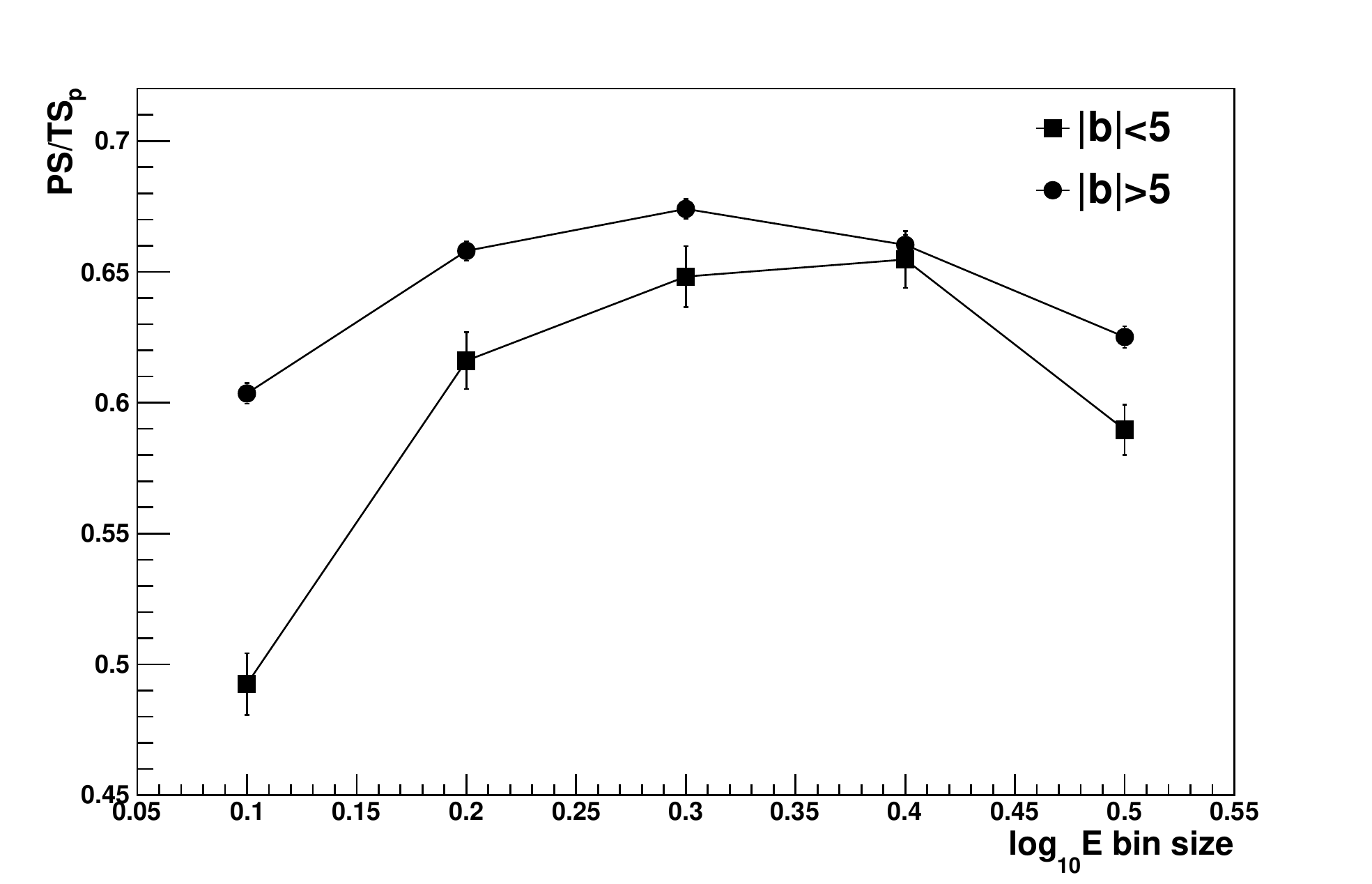}
  \caption{Position of the maximum of the PS/TS distribution of low and high-latitude sources as a function of the energy binning.}
  \label{fig:ebinoptimization}
\end{figure}

So far we have used an energy binning of 0.1 in $\log_{10}{E}$. Using the optimized spatial selection, we compute the PS/TS ratio for a $\log_{10}{E}$ bin width of 0.2, 0.3, 0.4 and~0.5. Figure~\ref{fig:ebinoptimization} shows that increasing the $\log_{10}{E}$ bin width reduces the difference between low and high Galactic latitude sources and that a maximum PS/TS ratio of~0.65 is reached for a bin width of about 0.3.

We conclude that the PS sensitivity is on average about 65\% of the TS sensitivity. The sensitivity loss with respect to the TS is the cost of performing the energy dependent integration. We note that the loss is rather modest, considering the pixel count information that is lost by the energy dependent integration, and is somewhat mitigated by the gain in computation speed.

Using a larger energy binning has also the advantage of slightly increasing the fraction of the spectral bins that are in the Gaussian regime, and for which the absolute meaning of the systematic uncertainty is fully taken into account by the weights, as explained in Section~\ref{sec:wLL}. Since we use systematic uncertainties of about 3\%, they do not have any impact on the bins with very few counts. The fraction is thus computed with respect to the number of spectral bins with at least one predicted count. For the RoI centered on the North Galactic pole, the average of this fraction over the pixels increases from 44\% to 50\% when the energy binning in $\log_{10}{E}$ is widened from 0.1 to 0.3. For the RoI with much more statistics centered on the Galactic center, it increases from 60\% to 65\%.

\subsection{Extended deviation case}

Since most of the 4FGL-DR2 sources are point-like sources, the optimized spatial selection, especially $p_2 = 0.1\degr$, is optimal for point-like sources. As noted in Section~\ref{sec:integratedcountspectra}, we expect that a larger value of $p_{2}$ increases the PS sensitivity to extended deviations. In order to investigate this possibility, we compute the PS as a function of $p_2$ for a simulated source whose spatial extension is a uniform disk with a $0.5\degr$ radius and also for a point-like source. To perform this test, we use the catXcheck RoI centered on $(l=0\degr,b=20\degr)$, which has no 4FGL-DR2 source within $3\degr$ of its center. We place the simulated source at the center of the RoI and set its spectrum to a power law with a photon index of 2. In both the point-like and extended cases, the flux is chosen such that the resulting PS is about 30 on average.

\begin{figure}[ht]
  \centering
  \includegraphics[width=9.5cm]{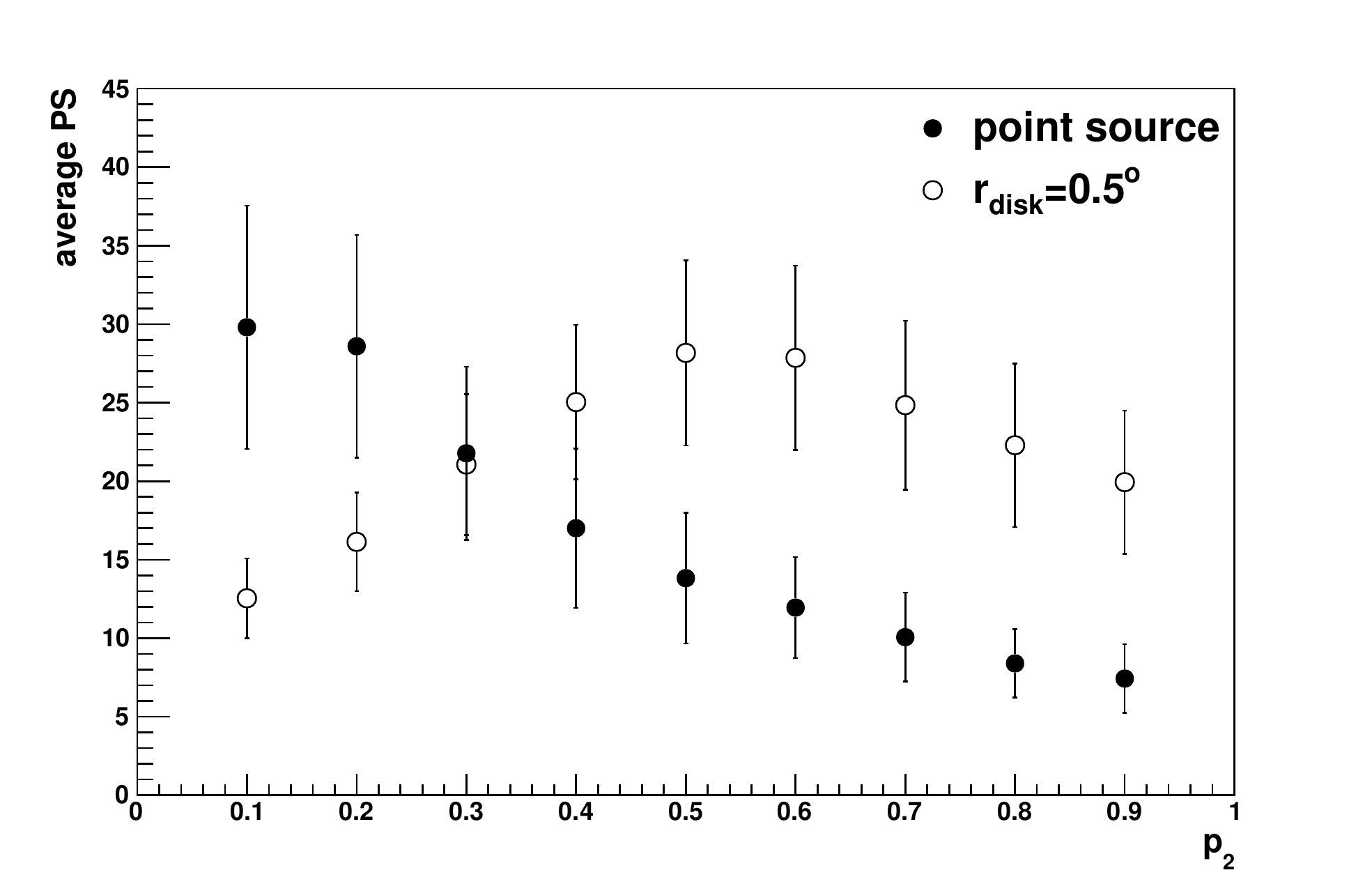}
  \caption{Variation of the average PS with $p_2$ for a point-like source (filled circle) and an extended source (empty circle), whose spatial model is a uniform disk with a $0.5\degr$ radius. The average is computed over 20 simulations and the error bars correspond to the root mean square.}
  \label{fig:SimuSource_p3scan}
\end{figure}

We simulate 20 mock 3D count data maps with the simulated source in the model and compute the PS map with these data maps and the predicted count map when the simulated source is removed from the model. We then compute the average and the root mean square of the PS map maximum over the 20 simulations. Figure~\ref{fig:SimuSource_p3scan} shows the variation of the average PS with $p_2$. In the point-like case, the maximum is obtained with $p_2 = 0.1\degr$, as expected. In the case of the extended source, the maximum is reached for $p_2$ between 0.5 and $0.6\degr$, which is of the order of the source extension, and is about twice the PS measured with $p_2=0.1\degr$, confirming that the PS sensitivity to extended deviations is significantly enhanced by increasing $p_2$.

\begin{figure}[ht]
  \centering
  \includegraphics[width=9.3cm]{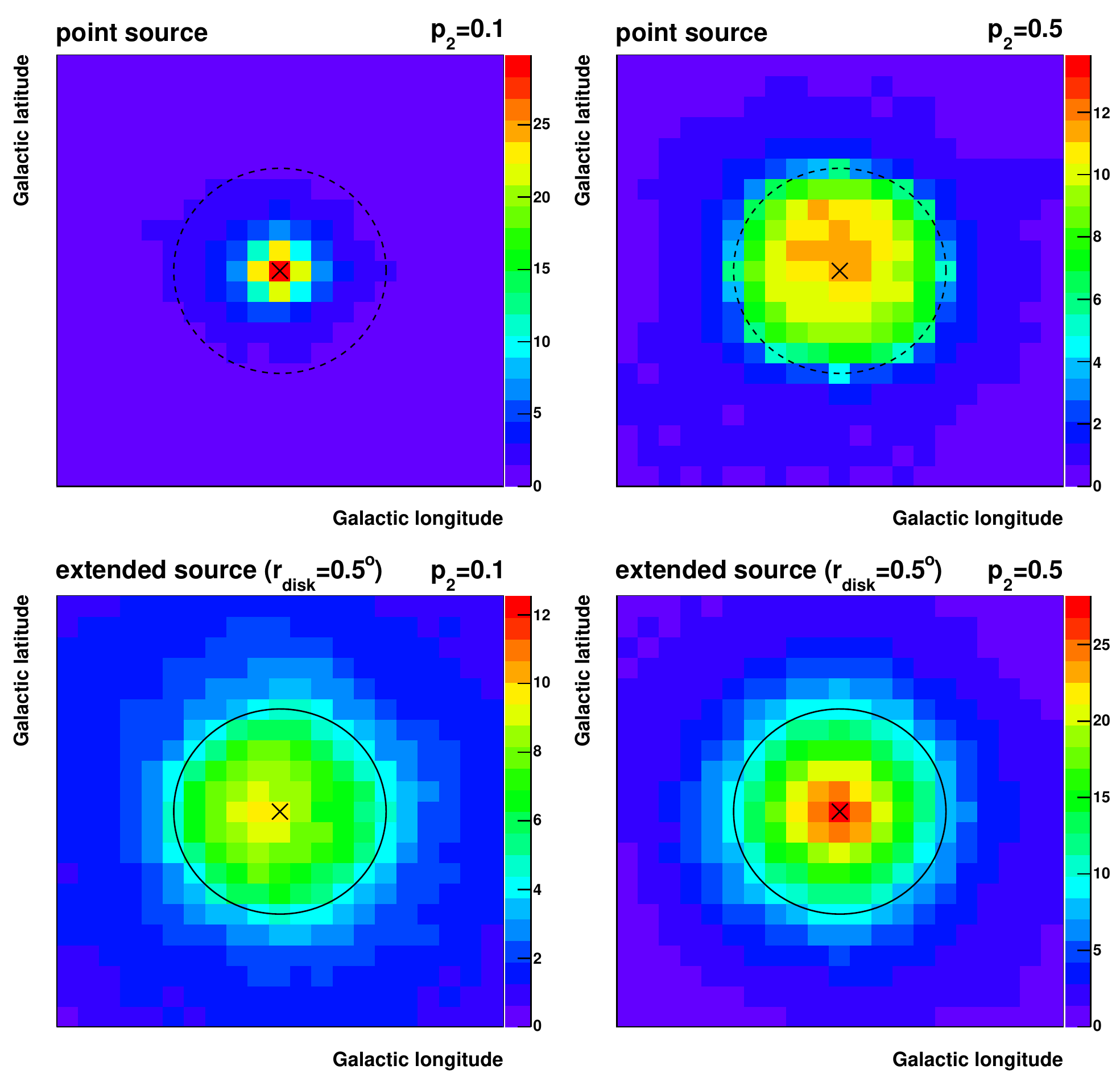}
  \caption{$2\degr \times 2\degr$ region of the average PS map around the simulated source for a point-like source (top) and an extended source (bottom), whose spatial model is a uniform disk with a $0.5\degr$ radius. The PS is computed with $p_2=0.1$ (left) and 0.5 (right). The pixel size is $0.1\degr$ and the radius of the superimposed circle is $0.5\degr$. The cross indicates the position of the simulated source.}
  \label{fig:SimuSource_comp_pointsource_disk}
\end{figure}

Figure~\ref{fig:SimuSource_comp_pointsource_disk} shows some of the average PS maps obtained with these simulations. For the point-like source, the PS map with $p_2=0.1\degr$ is clearly peaked at the position of the source, whereas with $p_2=0.5\degr$ it is almost flat over a disk of radius $\sim 0.5\degr$. The results are reversed in the case of the extended source: the average PS map is flatter with $p_2=0.1\degr$ than with $p_2=0.5\degr$. But with the latter, the PS map is not as peaked as in the point-like source case with $p_2=0.1\degr$. We conclude that a flat PS peak with $p_2=0.1\degr$ likely corresponds to an extended deviation, which can be further confirmed by computing the PS with a larger $p_2$. We note however that the PS statistical fluctuations, visible in Figure~\ref{fig:SimuSource_p3scan}, are such that a precise extension measurement requires performing a standard TS-based analysis.

\subsection{PS calibration}

For each of the 438 RoIs, catXcheck produces a PS map and we look for $|\mathrm{PS}|_\mathrm{max}$, the maximum PS measured in the RoI. The asymptotic behavior of the expected CCDF of $|\mathrm{PS}|_\mathrm{max}$ can be easily derived from that of the PS if there is no correlation between pixels: it corresponds to a $10^{-x}$ function scaled by the trial correction factor $120 \times 120$ or, equivalently, shifted horizontally by $\log_{10}(120 \times 120) \sim 4.16$. In order to check the existence of correlations, we simulate 100000 mock 3D count data maps and compute the corresponding PS maps, with the spatial selection parameters $p_0=4$ and $p_1 = 0.9$ and a $\log_{10}{E}$ bin width of 0.1. No systematic effect is included in this simulation and all weights are thus set to 1. We perform this test for two RoIs: the ones centered on the Galactic center and on the North Galactic pole. These two RoIs allow us to test very different situations in terms of statistics, as is clear from the integrated spectra at their respective centers shown in Figure~\ref{fig:integratedcountspectra}.

\begin{figure}[ht]
  \centering
  \includegraphics[width=9.5cm]{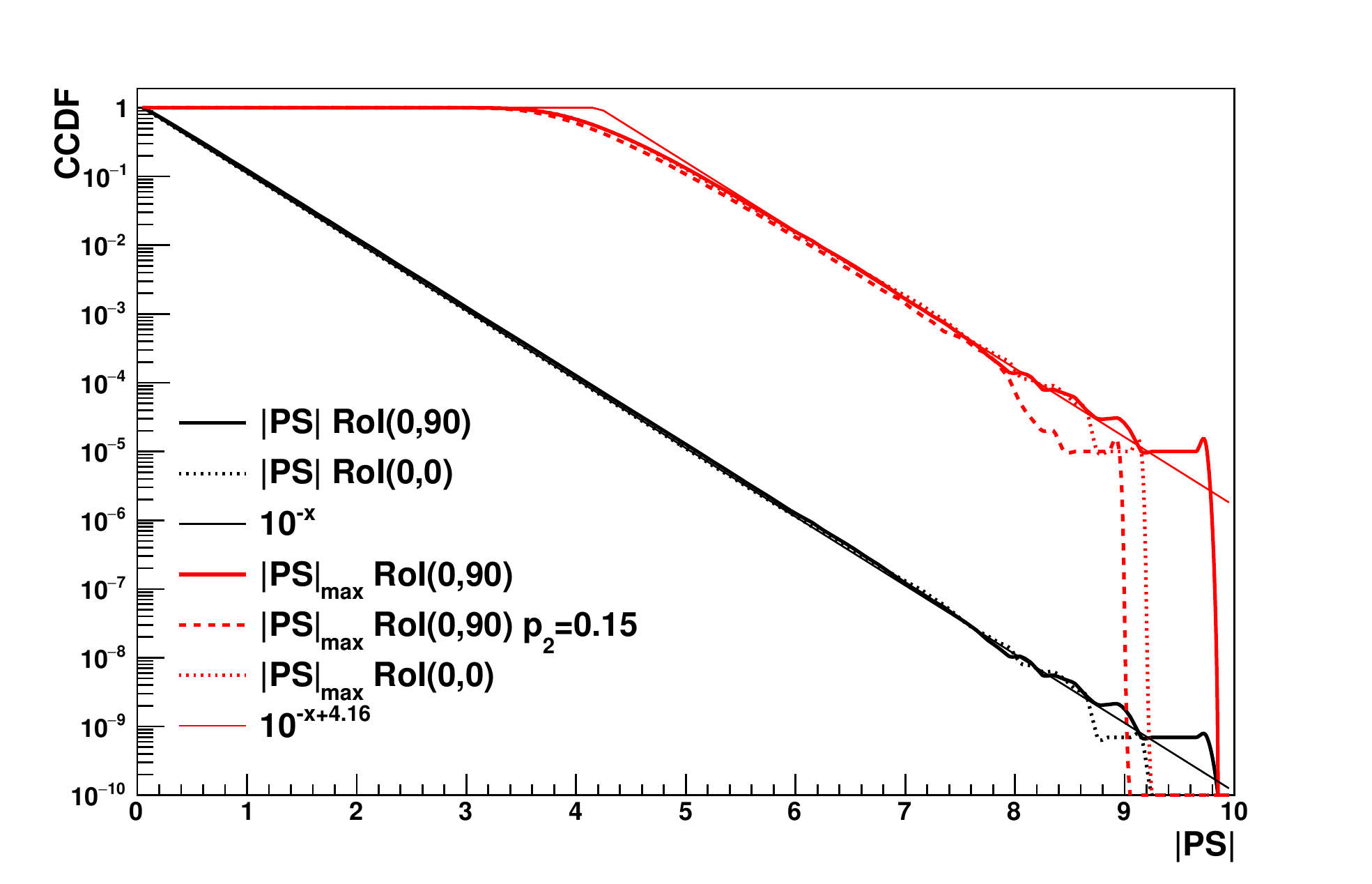}
  \caption{Simulations of the Galactic North pole (solid) and the Galactic center (dotted) RoIs: $|\mathrm{PS}|$ (black) and $|\mathrm{PS}|_\mathrm{max}$ (red) CCDF as well as the corresponding expected distributions (thin solid).}
  \label{fig:PScalibrationcumul}
\end{figure}

We first check the $|\mathrm{PS}|$ CCDF derived from all the pixels, which is displayed in Figure~\ref{fig:PScalibrationcumul} for PS computed with $p_2=0.10\degr$. For the two RoIs, it closely follows the $10^{-x}$ expectation: the deviation from expectation expressed as an error on PS is within 3\%. Figure~\ref{fig:PScalibrationcumul} also shows the $|\mathrm{PS}|_\mathrm{max}$ CCDF. There is no significant deviation at large $|\mathrm{PS}|_\mathrm{max}$ values from the expectation after correction for the number of pixels. This demonstrates that the level of correlation between pixels is very low. When the PS is computed with $p_2=0.15\degr$, the $|\mathrm{PS}|$ CCDF is almost identical to $p_2=0.10$ but the $|\mathrm{PS}|_\mathrm{max}$ CCDF is systematically $\sim3$\% below the expectation, as shown in Figure~\ref{fig:PScalibrationcumul}. It actually follows the expectation with a trial correction corresponding to an effective number of pixels equal to $\sim 110^2$, smaller than $120^2$, expected when there is no correlation. This shows that, for a pixel size of $0.1\degr$, there is some correlation between adjacent pixels with $p_2=0.15\degr$.

\begin{figure}[ht]
  \centering
  \includegraphics[width=9.5cm]{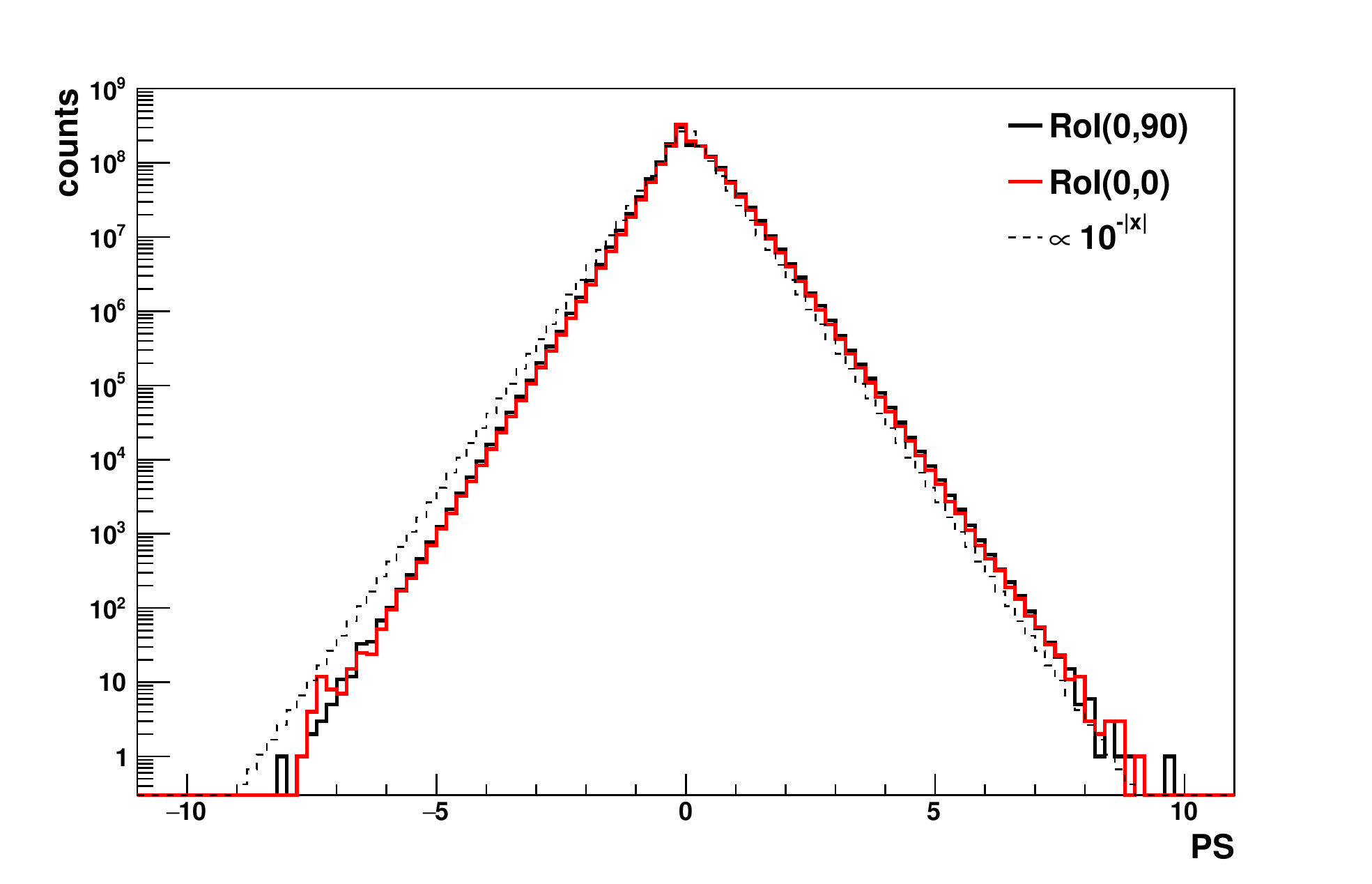}
  \caption{The distribution of PS in simulations of the Galactic North pole (solid) and the Galactic center (dotted) RoIs, as well as the corresponding expected distributions (thin solid).}
  \label{fig:PScalibrationsigned}
\end{figure}

The distribution of the signed PS differs from the exponential expectation in two aspects, as can be seen in Figure~\ref{fig:PScalibrationsigned} for the two RoIs. The first difference is a systematic $\sim 0.1$ horizontal shift with respect to the expectation. This is very likely due to the simple prescription we use to derive the PS sign. The second difference is at the peak around 0 and is the consequence of a lack of precision for p-values close to 1. This feature could be reduced by changing the $L$ pdf computation parameters (especially $n_\mathrm{pdf}$). However, the large p-value region is not critical when looking for significant data/model deviations so it does not seem useful to slow down the PS computation only to reach a better agreement with the expectation around 0. We note that, when PS is expressed in $\sigma$ units, the notch at 0 is even more apparent.

\section{Results of the 4FGL-DR2 verification} \label{sec:catXcheck}

In this Section we present the results of the catXcheck analysis on the 4FGL-DR2 catalog. The PS map of each RoI is computed with the optimized spatial selection parameters and a $\log_{10}{E}$ bin width of 0.3 (see Appendix~\ref{app:PSproductionsteps} for a detailed description of the PS map production). Since there are $120 \times 120$ pixels in each PS map and we consider 438 RoIs, the $3,4$ and $5\sigma$ thresholds correspond to $\mathrm{PS}=9.4, 11$ and $13$, respectively. In the following we use $\mathrm{PS}>11$ to select significant deviations.

\begin{figure}[ht]
  \centering
  \includegraphics[width=9.5cm]{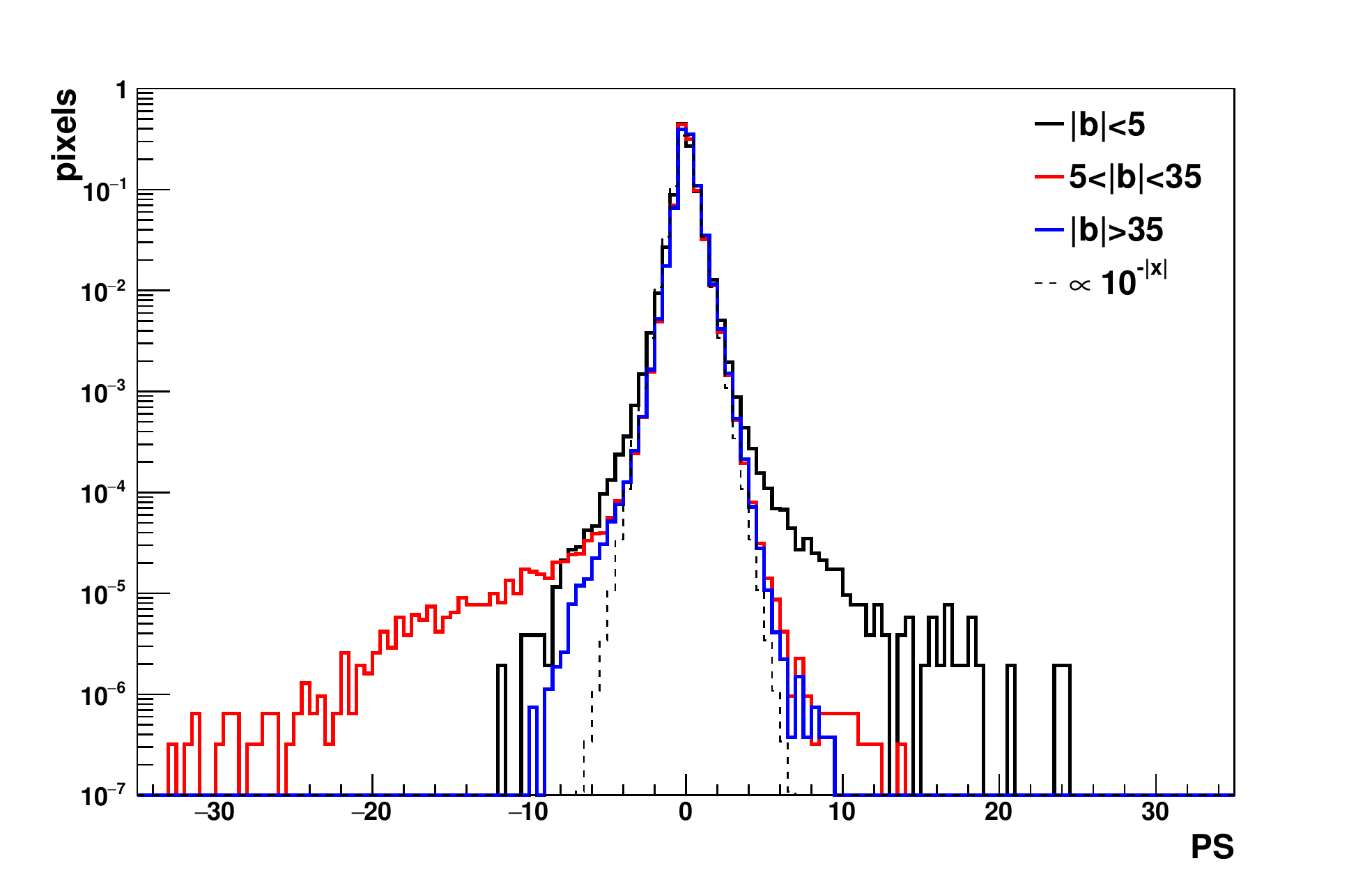}
  \caption{PS distribution obtained in the verification of the 4FGL-DR2 catalog (10~years of data) for the three Galactic latitude selections. The thin solid histogram corresponds to the expected $10^{-|x|}$ distribution. All histograms are normalized such that their integral is 1.}
  \label{fig:catXcheck_PSband}
\end{figure}

Figure~\ref{fig:catXcheck_PSband} shows the PS distributions for three Galactic latitude samples: low ($|b|<5\degr$), mid ($5\degr<|b|<35\degr$) and high ($|b|>35\degr$) latitudes. There is no significant deviation in the high-latitude selection whereas the low and mid-latitude distributions exhibit positive and negative broad tails, respectively.

\begin{figure}[ht]
  \centering
  \includegraphics[width=9.3cm]{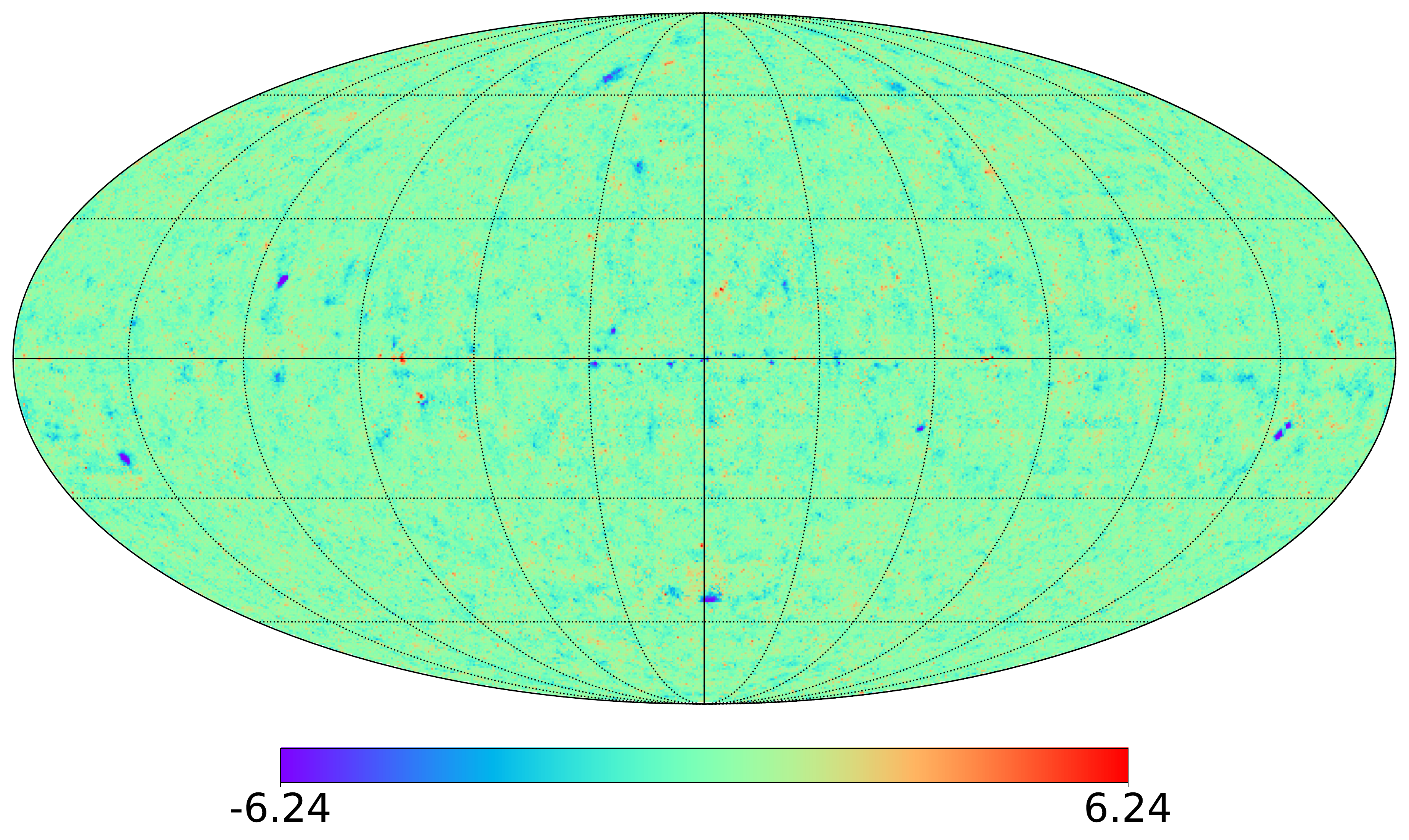}
  \caption{All-sky PS map in Galactic coordinates (Mollweide projection).}
  \label{fig:catXcheck_allskymap}
\end{figure}

From the PS maps of the 438 RoIs, it is possible to construct an all-sky PS map. We use a HEALPix~\citep{healpix} map in Galactic coordinates with $N_\mathrm{side}=256$ and we set the PS of each pixel to the maximum of the PS found among the individual RoI pixels falling into that HEALPixel. The resulting all-sky PS map is shown in Figure~\ref{fig:catXcheck_allskymap}.

The most significant negative deviations (clusters of many pixels with $\mathrm{PS}<-11$) correspond to five negative spots located at $(l,b) \sim (112.8\degr,16.6\degr)$, $(157.6\degr,-21.2\degr)$, $(302.6\degr,-14.7\degr)$, $(206.9\degr,-16.7\degr)$ and $(205.0\degr,-14.1\degr)$. Since they are all close to large molecular clouds (the first one is associated with Cepheus, the second with Perseus, the third with Chamaeleon and the two last ones with Orion~B), they are due to imperfections in the modelling of the Galactic diffuse emission\footnote{\url{https://fermi.gsfc.nasa.gov/ssc/data/analysis/software/aux/4fgl/Galactic_Diffuse_Emission_Model_for_the_4FGL_Catalog_Analysis.pdf}}. A more detailed analysis of these imperfections will be included in the forthcoming 4FGL-DR3 catalog publication.

\begin{figure}[ht]
  \centering
  \includegraphics[width=9.3cm]{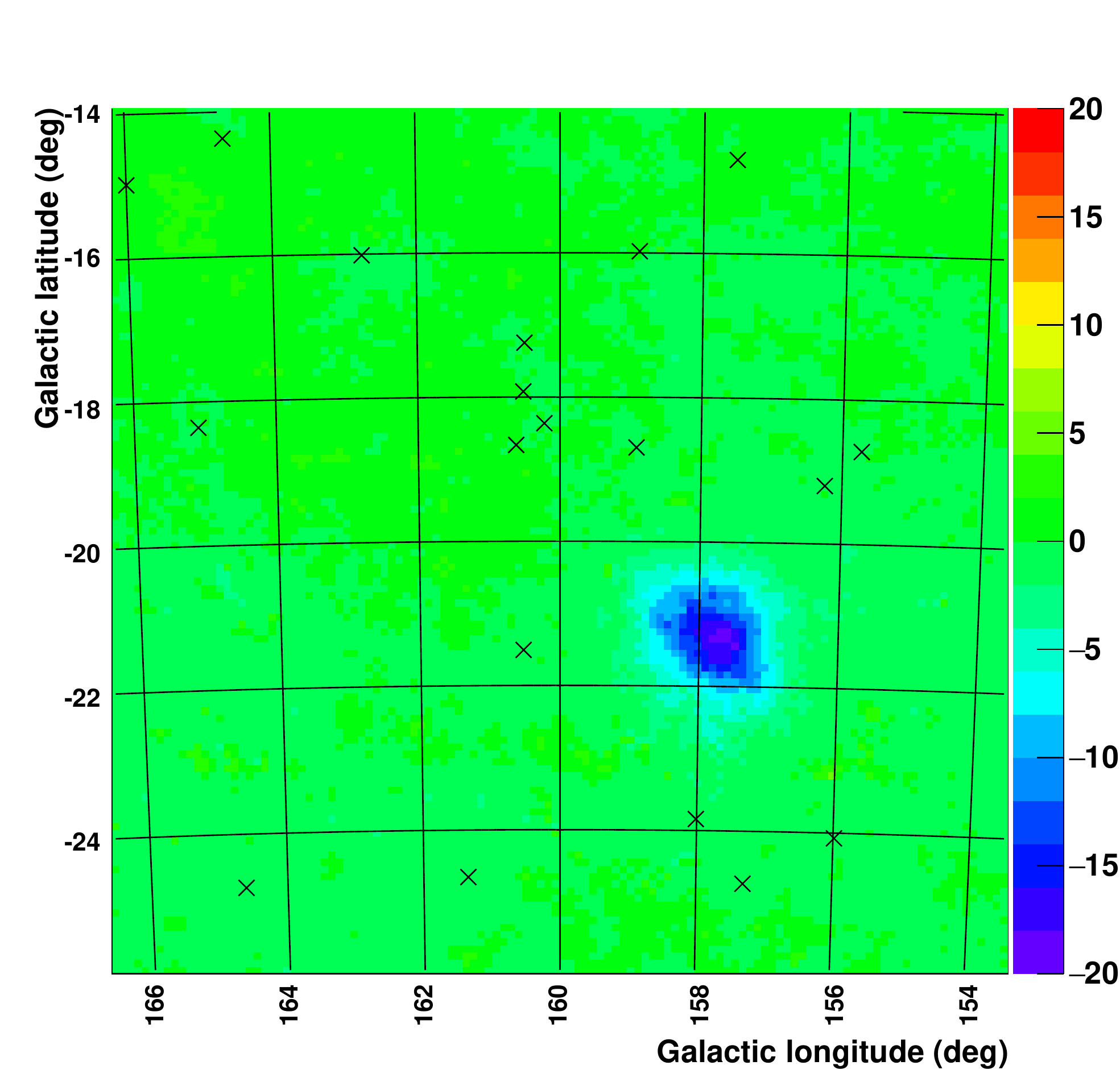}
  \caption{PS map of the RoI centered on $(l=160\degr,b=-20\degr)$. The black crosses correspond to 4FGL-DR2 sources.}
  \label{fig:catXcheck_roi_-20_160_psmap}
\end{figure}

\begin{figure}[ht]
  \centering
  \includegraphics[width=9.5cm]{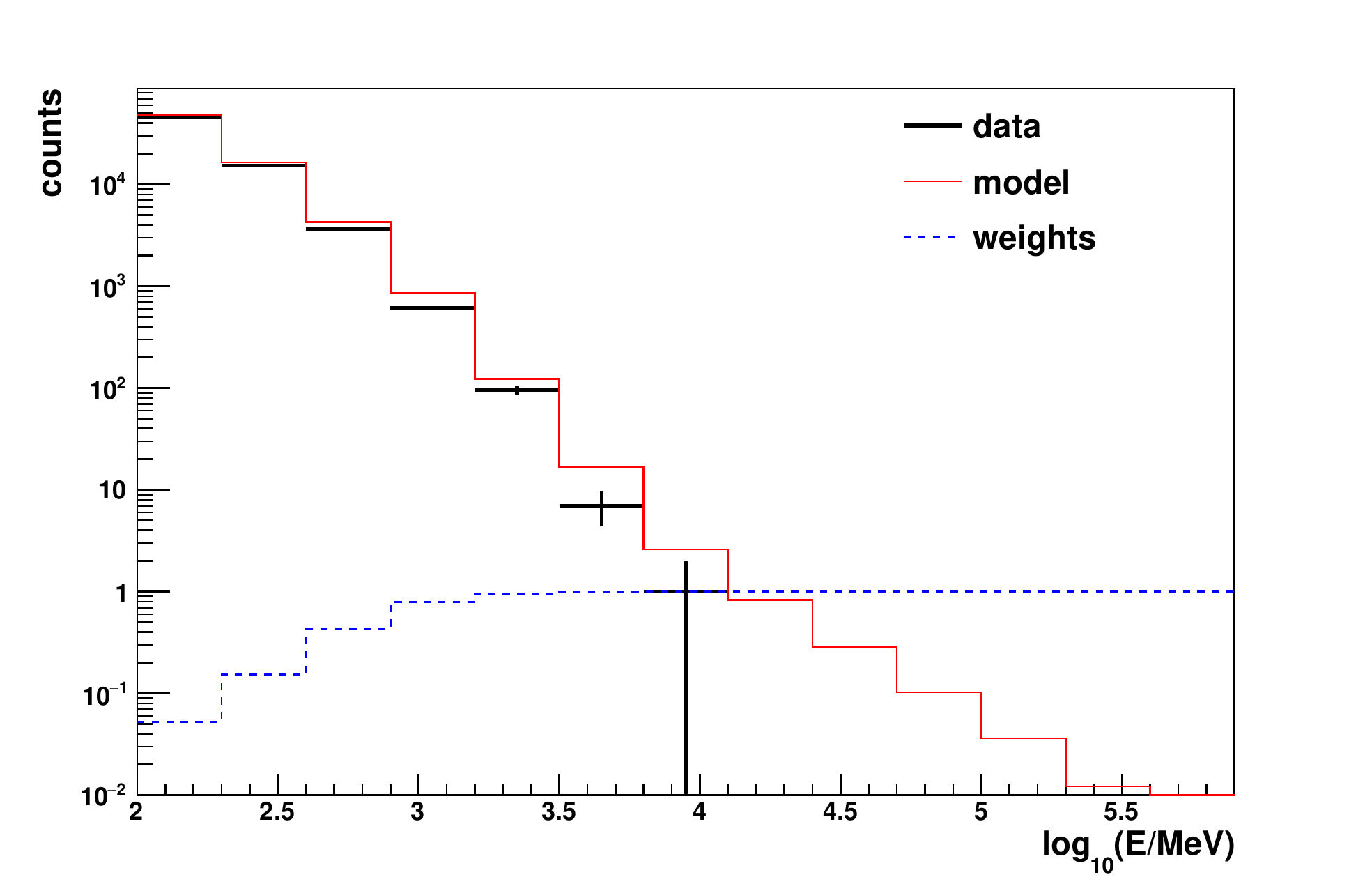}
  \caption{Integrated count spectra corresponding to the maximum PS in the RoI centered on $(l=160\degr,b=-20\degr)$, for data (black) and model (red), as well as the average likelihood weights (dashed blue).}
  \label{fig:catXcheck_roi_-20_160_spec}
\end{figure}

The PS map of the RoI centered on $(l=160\degr,b=-20\degr)$ containing the negative spot at $(l=157.6\degr,b=-21.2\degr)$ is shown in Figure~\ref{fig:catXcheck_roi_-20_160_psmap}. The data and model integrated count spectra corresponding to the pixel with the lowest PS ($-19.5$) are shown in Figure~\ref{fig:catXcheck_roi_-20_160_spec}. One can see that the model overpredicts the data in the energy range between 0.3 and 10~GeV. Since the PS is relatively flat around the minimum, we perform a PS scan over the spatial selection parameter $p_2$. A minimum of $-47.4$, much lower than $-19.5$, is obtained around $p_2=1.3\degr$, indicating that the deviation is extended.

\begin{figure}[ht]
  \centering
  \includegraphics[width=9.3cm]{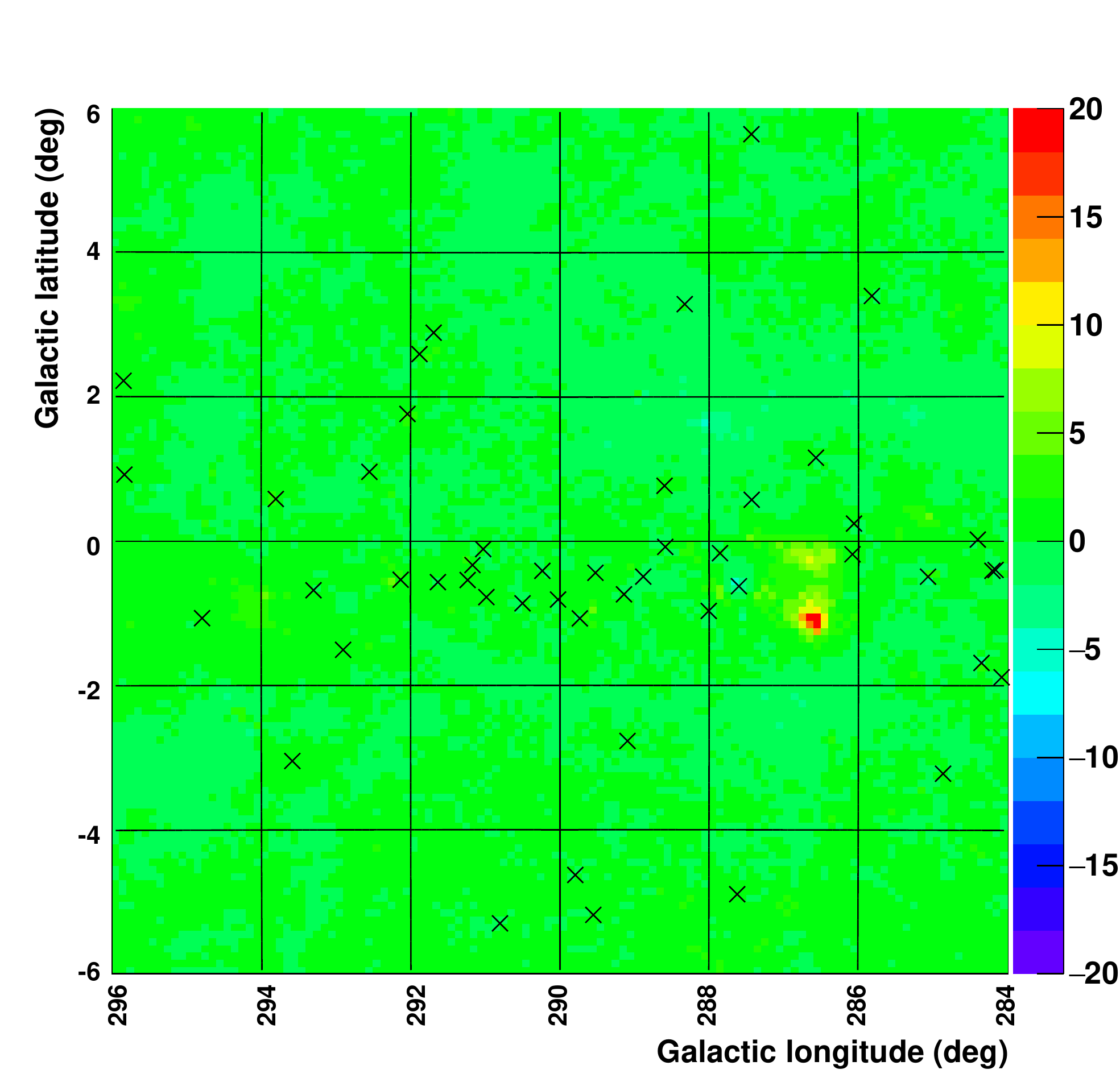}
  \caption{PS map of the RoI centered on $(l=290\degr,b=0\degr)$. The black crosses correspond to 4FGL-DR2 sources.}
  \label{fig:catXcheck_roi_0_290_psmap}
\end{figure}

\begin{figure}[ht]
  \centering
  \includegraphics[width=9.5cm]{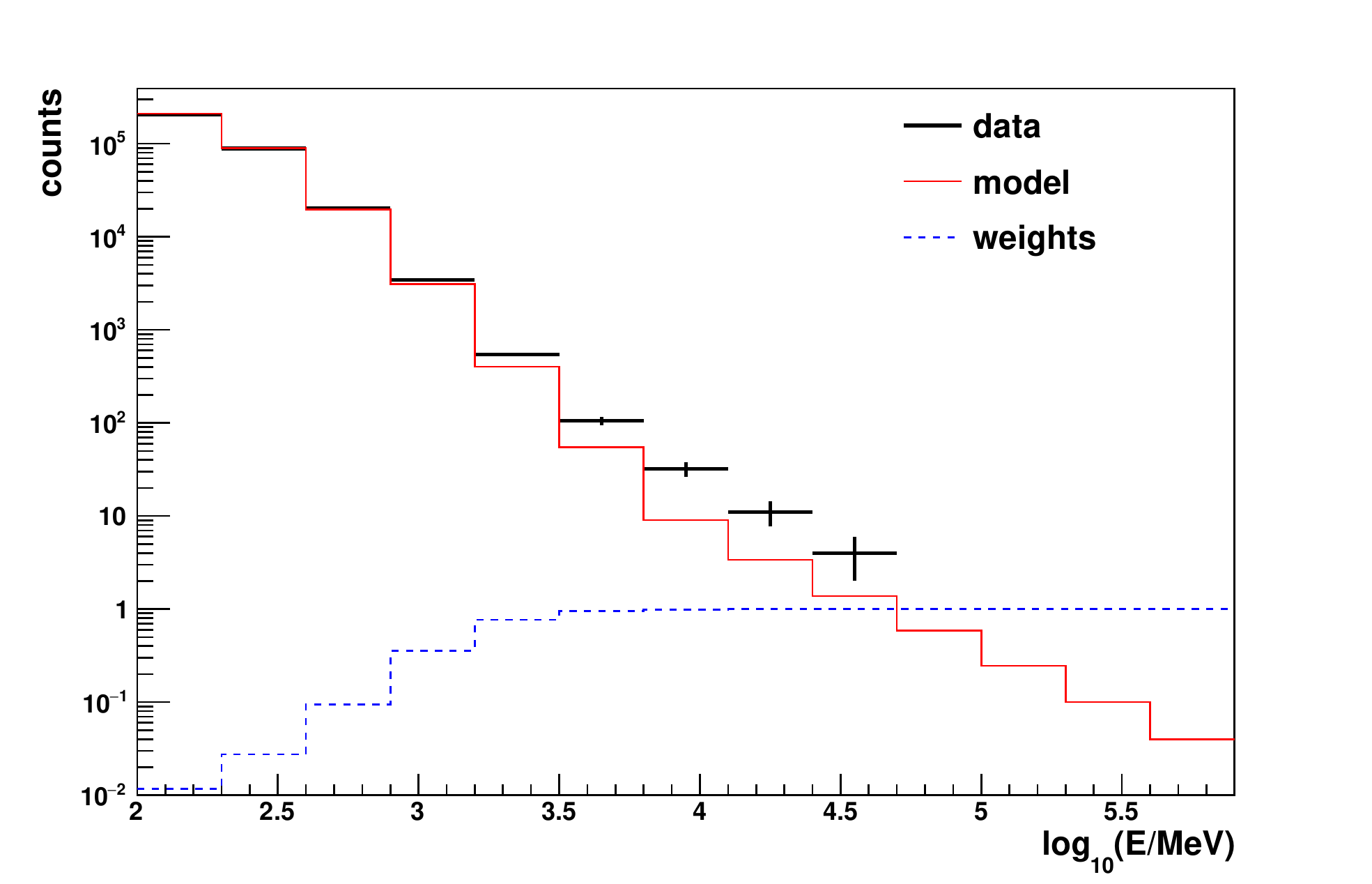}
  \caption{Integrated count spectra corresponding to the maximum PS in the RoI centered on $(l=290\degr,b=0\degr)$, for data (black) and model (red), as well as the average likelihood weights (dashed blue).}
  \label{fig:catXcheck_roi_0_290_spec}
\end{figure}

Both Figure~\ref{fig:catXcheck_PSband} and Figure~\ref{fig:catXcheck_allskymap} show that there are several significant positive deviations. The one in the mid-latitude selection (PS=12.3) is located in the RoI centered on $(l=110\degr,b=-30\degr)$ and it corresponds to a point-like excess which will be included in the next 4FGL-DR3 catalog~({\it Fermi}-LAT Collaboration, in preparation). It is also the case for most of the deviations in the low-latitude selection. The most significant one (PS=24.25) is located at $(l=286.55\degr,b=-1.15\degr)$. The PS map of the corresponding RoI is shown in Figure~\ref{fig:catXcheck_roi_0_290_psmap} and the data and model integrated count spectra at the maximum PS position are shown in Figure~\ref{fig:catXcheck_roi_0_290_spec}. The excess of counts in data is visible above 1~GeV, which is typical of a missing source in the model. After investigation, it was found that this excess corresponds to the bright optical Nova ASASSN-18fv~\citep{NOVA_ATel_optical} at $(l=286.573\degr,b=-1.088\degr)$, whose gamma-ray emission has been detected by {\it Fermi}-LAT~\citep{NOVA_ATel_fermi,NOVA_paper} around April 14, 2018. This source will be included in the 4FGL-DR3 catalog.

\begin{figure}[ht]
  \centering
  \includegraphics[width=9.5cm]{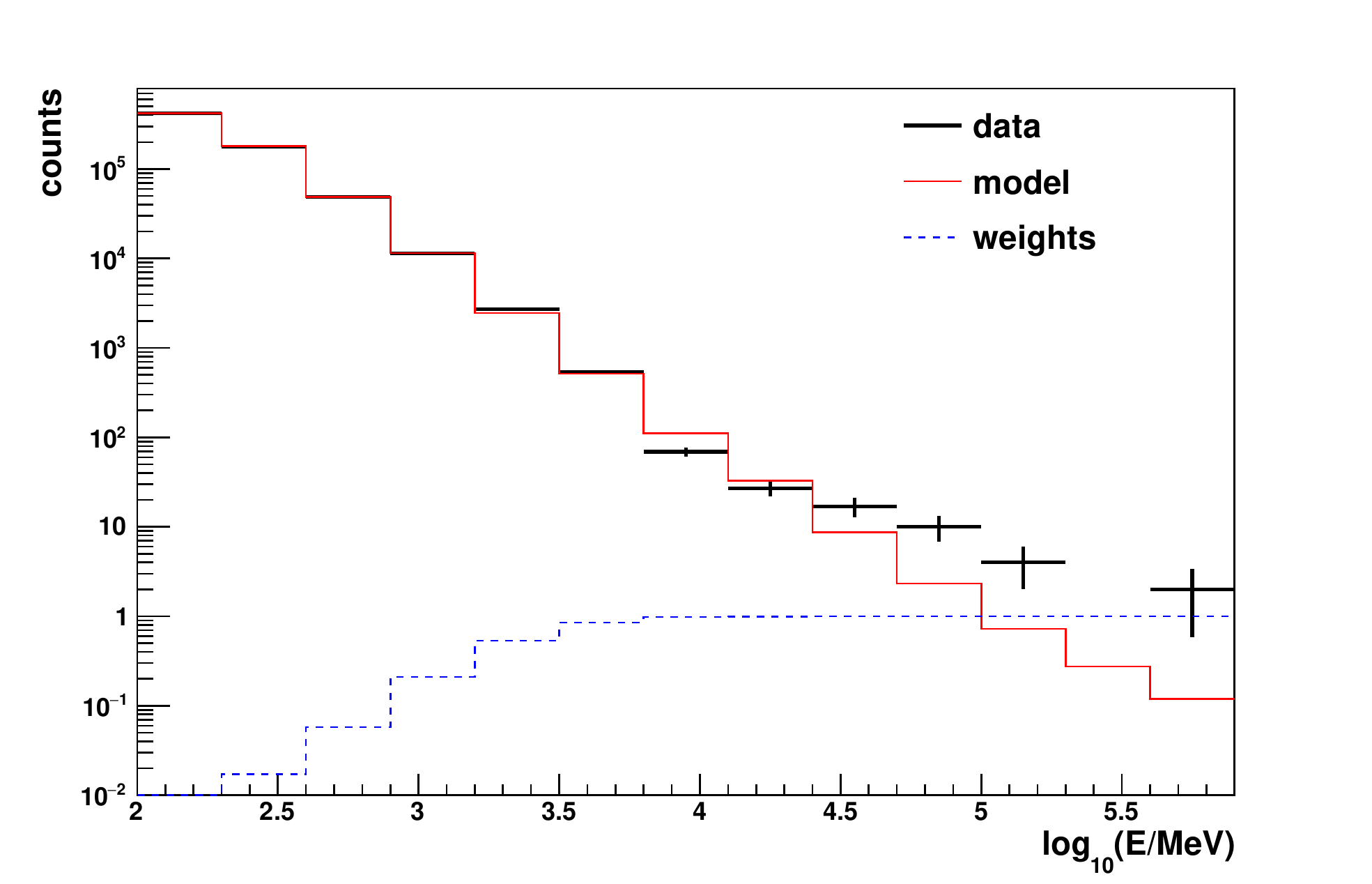}
  \caption{Integrated count spectra at the position of the gamma-ray binary LS~5039, for data (black) and model (red), as well as the average likelihood weights (dashed blue).}
  \label{fig:catXcheck_roi_0_20_spec}
\end{figure}

The RoI centered on $(l=20\degr,b=0\degr)$ provides an example of a source whose spectrum is mismodelled. The maximum of the PS map is $9.28$ and its position corresponds to the gamma-ray binary LS~5039~\citep{LS5039_2009,LS5039_2012,LS5039_2016}. The 4FGL-DR2 analysis considers three spectral models (power law, power law with subexponential cutoff and log normal) for each source and selects the best model. The LS~5039 spectrum is modelled with a log normal but this spectral shape is not able to reproduce the integrated count spectrum, as can be seen in Figure~\ref{fig:catXcheck_roi_0_20_spec}. This is the consequence of the presence of a second spectral component, first reported by~\citet{LS5039_2012} and recently confirmed by~\citet{LS5039_2021}.

\section{Conclusion} \label{sec:conclusion}

{\it Fermi}-LAT analyses are generally based on a binned log-likelihood fit of a 3D count map but there is no fast, reliable and sensitive tool available to check the goodness-of-fit (including the case of negative residuals). In order to overcome the lack of such a tool, we have developed a new method that allows {\it Fermi}-LAT data users to quantify efficiently the data/model agreement after performing a fit of an RoI. The method is based on integrating the observed and predicted counts over an energy dependent region around each pixel of the map and on computing a deviation estimator, named PS, between the integrated data and model count spectra. This method can incorporate the likelihood weights that are used in {\it Fermi}-LAT analyses to take into account some systematic uncertainty.

In order to minimize the computation time, the PS algorithm has been optimized while ensuring a PS precision of 3\%. The PS statistical calibration has been checked with simulations and its average sensitivity to a point-like source deviation has been measured at 65\% of the TS. This lower sensitivity is naturally explained by the energy dependent spatial integration which dilutes the 3D information into 1D count spectra. But this integration allows the PS map computation\footnote{A python script computing PS maps is available at the User contribution page of the Fermi Science Support Center web site: \url{https://fermi.gsfc.nasa.gov/ssc/data/analysis/user/}. The links to the script and the documentation are \url{https://fermi.gsfc.nasa.gov/ssc/data/analysis/user/gtpsmap/gtpsmap.py} and \url{https://fermi.gsfc.nasa.gov/ssc/data/analysis/user/gtpsmap/README}.} to be much faster than for TS maps. This is very convenient while optimizing the sky model of an RoI. Another important advantage of the PS is that it is sensitive to both positive and negative deviations.

The use of PS maps to check 4FGL-DR2, the latest of the {\it Fermi}-LAT general catalog, has proven to be useful, reporting some positive deviations, actually corresponding to gamma-ray sources, and some negative deviations, related to imperfections in the modelling of the Galactic diffuse emission. The same verification is being performed in the preparation of 4FGL-DR3, the next catalog based on 12~years of data.

\begin{acknowledgements}
We thank our {\it Fermi}-LAT collaborators Jean Ballet and Matthew Kerr for fruitful discussions.
 
The \textit{Fermi} LAT Collaboration acknowledges generous ongoing support
from a number of agencies and institutes that have supported both the
development and the operation of the LAT as well as scientific data analysis.
These include the National Aeronautics and Space Administration and the
Department of Energy in the United States, the Commissariat \`a l'Energie Atomique
and the Centre National de la Recherche Scientifique / Institut National de Physique
Nucl\'eaire et de Physique des Particules in France, the Agenzia Spaziale Italiana
and the Istituto Nazionale di Fisica Nucleare in Italy, the Ministry of Education,
Culture, Sports, Science and Technology (MEXT), High Energy Accelerator Research
Organization (KEK) and Japan Aerospace Exploration Agency (JAXA) in Japan, and
the K.~A.~Wallenberg Foundation, the Swedish Research Council and the
Swedish National Space Board in Sweden.
 
Additional support for science analysis during the operations phase is gratefully
acknowledged from the Istituto Nazionale di Astrofisica in Italy and the Centre
National d'\'Etudes Spatiales in France. This work performed in part under DOE
Contract DE-AC02-76SF00515.
\end{acknowledgements}

\begin{appendix}

\section{Iterative computation of the log-likelihood pdf} \label{app:iterativecomputation}

In this Appendix, we describe the iterative procedure to compute the probability distribution function (pdf) of the random variable $L$ defined in Equation~\ref{eq:loglike} of Section~\ref{sec:deviationprobability}. Let $N_\mathrm{bins}$ be the number of bins of the model integrated count spectrum and $m_i$ the model integrated counts in bin $i$. Let $\mathrm{pdf}_i$ be the probability distribution function corresponding to the first $i$ bins. $\mathrm{pdf}_1$, corresponding to the first bin, is simply given by the Poisson distribution.

In order to build $\mathrm{pdf}_i$ from $\mathrm{pdf}_{i-1}$, we must consider all the ways that a realization of the data could alter the values of the log-likelihood. Given the model integrated counts $m_i$, drawing a value of the counts $k$ will change $L$ by $-\log \mathcal{P}(k,m_i)$.  Thus, the iteration consists of computing the probability weighted sum of all the possible values of $k$. In other words, each $k$ contribution corresponds to $L_{i-1}$ shifted by $-\log \mathcal{P}(k,m_i)$ and multiplied by $\mathcal{P}(k,m_i)$, and $\mathrm{pdf}_i$ is obtained by simply summing all these contributions:

\begin{equation} \label{eq:pdfiteration}
\mathrm{pdf}_{i}(x) = \sum_{k} \mathcal{P}(k,m_i) \times \mathrm{pdf}_{i-1} \left( x+\log \mathcal{P}(k,m_i) \right)
\end{equation}

In order to minimize the computing time, we apply the following prescriptions:
\begin{itemize}
\item the $L$ pdf is computed as an histogram with $n_\mathrm{pdf} \times N_\mathrm{bins}$ bins on the range $[0,20N_\mathrm{bins}]$;
\item we introduce the precision parameter $\epsilon$ and, at each iteration, we only consider $k$ in the interval $[k_0(m_i),k_1(m_i)]$ defined as the narrowest interval such that $\sum_{k} \mathcal{P}(k,m_i) \geq 1-\epsilon$. The $k_0(m)$ and $k_1(m)$ parameterizations are given in~Appendix~\ref{app:kinterval}.
\end{itemize}

\begin{figure}[ht]
  \centering
  \includegraphics[width=9.0cm]{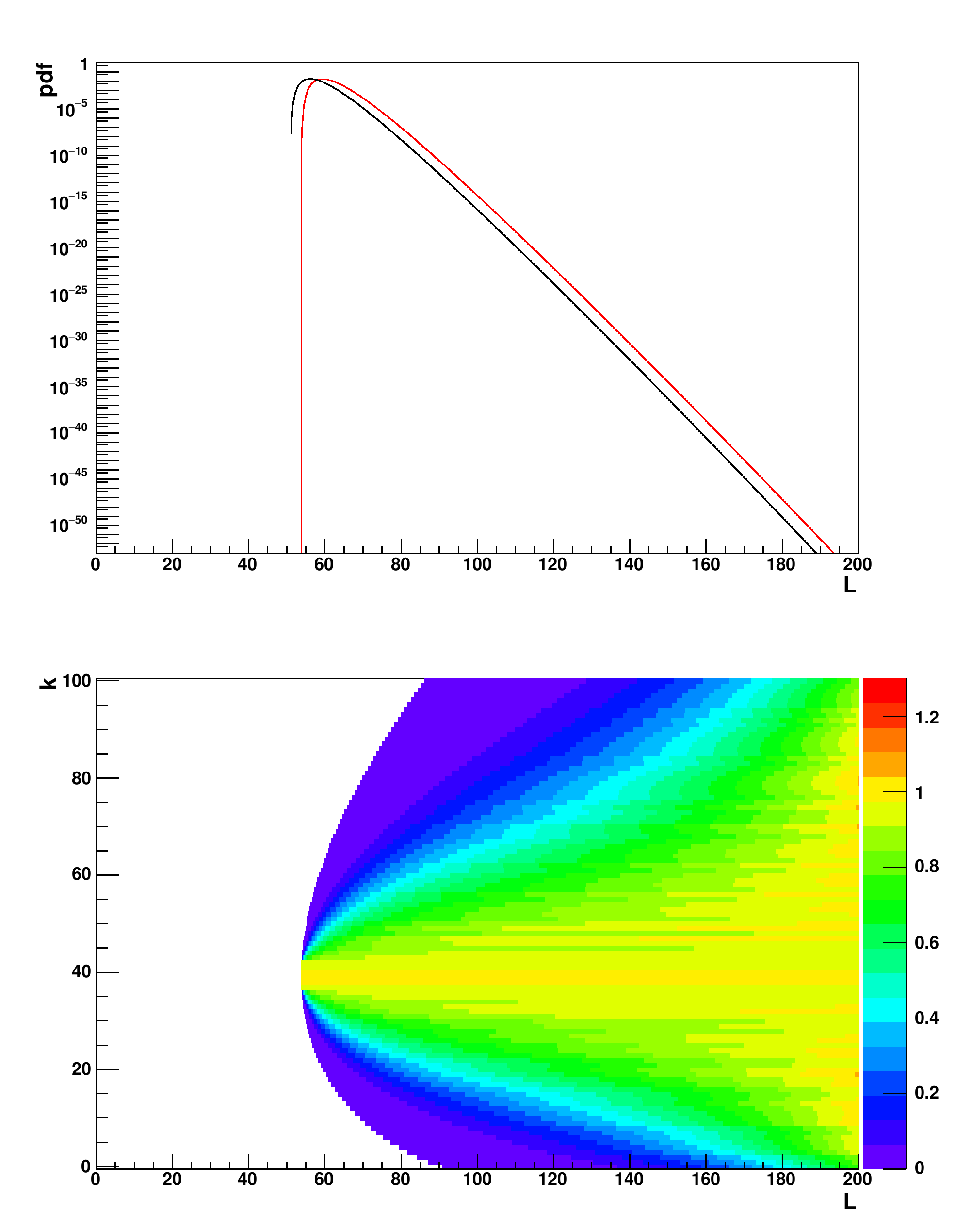}
  \caption{The $L$ distributions at the beginning (black) and the end (red) of an iteration of the $L$ pdf computation algorithm (top) and the corresponding individual $k$ contributions (bottom) after division by the maximum contribution (corresponding to $k=39$). The number of predicted counts in this step is 39.81 and the $y$-axis range corresponds to the $[k_0,k_1]$ interval obtained with $\epsilon = 10^{-15}$.}
  \label{fig:showiteration}
\end{figure}

An example of such an iteration is shown in Figure~\ref{fig:showiteration}. In this example, we use $n_\mathrm{pdf}=200$ and $\epsilon = 10^{-15}$ and the count spectrum corresponds to a 40~bins $E^{-2}$ spectrum between 100~MeV and 1~TeV with 10000 counts in the first bin. Figure~\ref{fig:showiteration} shows the 12th iteration, corresponding to the 12th spectral bin for which $m_i=39.81$. The $L$ pdf of the previous step is shown in black in the top panel. It is used to derive the $k$ contributions to the $L$ pdf for $k$ in $[0,100]$, as shown in the bottom panel after division by the maximum contribution (corresponding to $k=39$). Summing all these contributions along the $y$-axis leads to the $L$ pdf after the iteration that is shown in red in the top panel.

As explained in Section~\ref{sec:deviationprobability}, we first compute the contribution of the spectral bins with Gaussian statistics from the $\chi^2$ distribution with a number of degrees of freedom equal to the number of these bins, and then run the iterative computation over the spectral bins with Poisson statistics.

The algorithm to compute the $L$ pdf thus depends on three parameters:
\begin{itemize}
\item $\epsilon$ defining the interval $[k_0(m),k_1(m)]$;
\item $n_\mathrm{pdf}$ defining the number of bins of the $L$ pdf histogram ($n_\mathrm{pdf} \times N_\mathrm{bins}$);
\item $N_g$, the lower limit on the number of counts to decide whether a spectral bin is in the Gaussian regime.
\end{itemize}

The choice of $N_g$ is not critical because of the steepness of the count spectra. For a $\log_{10}{E}$ bin size of 0.1, the count ratio of neighbouring bins is typically $~2$. So increasing $N_g$ from {\it e.g.} 100 to 200 moves at most one spectral bin from the Gaussian regime to the Poisson regime, which is not enough to change significantly the final $L$ pdf.

\begin{figure}[ht]
  \centering
  \includegraphics[width=9.5cm]{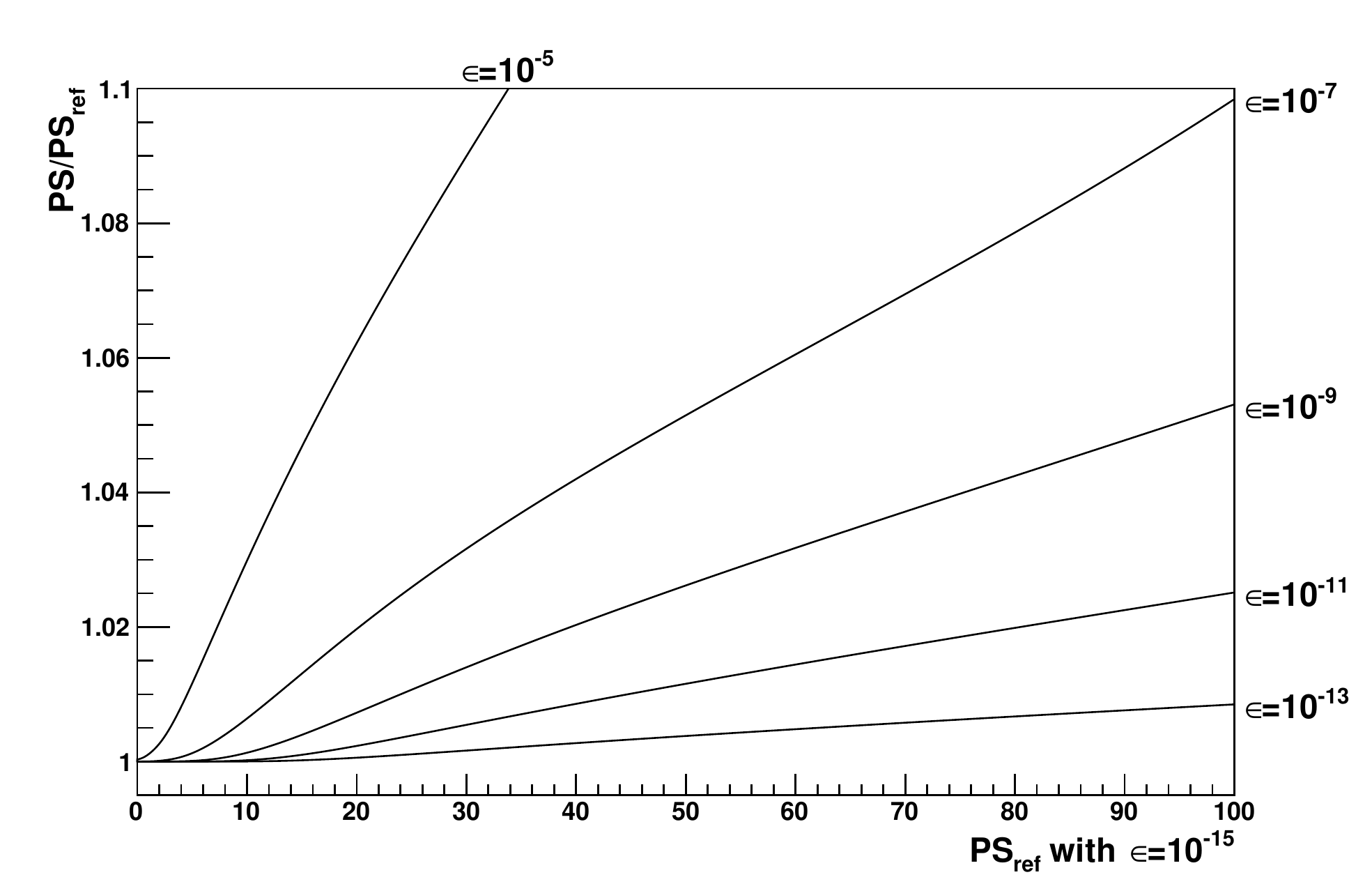}
  \caption{The ratio of PS to the reference PS as a function of the reference PS for several choices of $\epsilon$, the precision parameter that defines the interval of $k$ considered during the $L$ pdf computation. The ratio is actually evaluated as a function of $L$ and $L$ is converted to the PS for $\epsilon=10^{-15}$.}
  \label{fig:showPoissonprecision}
\end{figure}

In order to test the sensitivity to $\epsilon$, we set $n_\mathrm{pdf}=200$ (corresponding to a very fine binning of the $L$ pdf histogram) and $N_g=100$ and we compare the results obtained with $\epsilon$ ranging from $10^{-5}$ to $10^{-15}$, the latter being used as the reference. For each $\epsilon$ value, we compute the $L$ complementary cumulative distribution function (CCDF) in order to obtain the curve $\mathrm{PS}(L)$. Figure~\ref{fig:showPoissonprecision} shows the ratio of these curves as a function of the reference PS. One can see that $\epsilon=10^{-7}$ is small enough to get a 2\% precision up to $\mathrm{PS}=20$, that is to say well above $10$, which corresponds to $\sim 5\sigma$ for a typical $100 \times 100$ pixels map. Ensuring a 1\% precision up to $\mathrm{PS}=100$ can be useful when using the PS map around a positive excess to estimate the position of a possible point source at the origin of the deviation. In that case a smaller $\epsilon$ is needed, at the expense of computation time.

Regarding the choice of $n_\mathrm{pdf}$, we compare the results obtained with $n_\mathrm{pdf}=10$, 20, 50, 100 and 200, the latter being used as the reference. Differences larger than 1\% are only seen for $n_\mathrm{pdf}=10$ and 20.

In order to minimize computation time while ensuring a 2\% precision, we use the following parameters: $\epsilon=10^{-7}$, $n_\mathrm{pdf}=50$ and $N_g = 100$.

\section{$[k_\mathrm{min},k_\mathrm{max}]$ interval parameterization} \label{app:kinterval}

For a given number of predicted counts $m$ and a given precision parameter $\epsilon \ll 1$, let $k_\mathrm{min}$ and $k_\mathrm{max}$ be the boundaries of the narrowest $k$ interval such that $\sum_{k} \mathcal{P}(k,m) \geq 1-\epsilon$. In this Appendix, we derive the parameterizations of $k_\mathrm{min}$ and $k_\mathrm{max}$ as functions of $m$ and $\epsilon$. We note that the goal of the introduction of $\epsilon$ is mostly to be able to vary the level of precision of the $L$ pdf computation (described in Section~\ref{sec:deviationprobability}) in order to find an optimal choice with respect to the computation speed. As a consequence, the fact that $\sum_{k} \mathcal{P}(k,m)$ monotonically increases when $\epsilon$ decreases is more important than ensuring a high level of accuracy for these parameterizations.

\begin{figure}[ht]
  \centering
  \includegraphics[width=9.5cm]{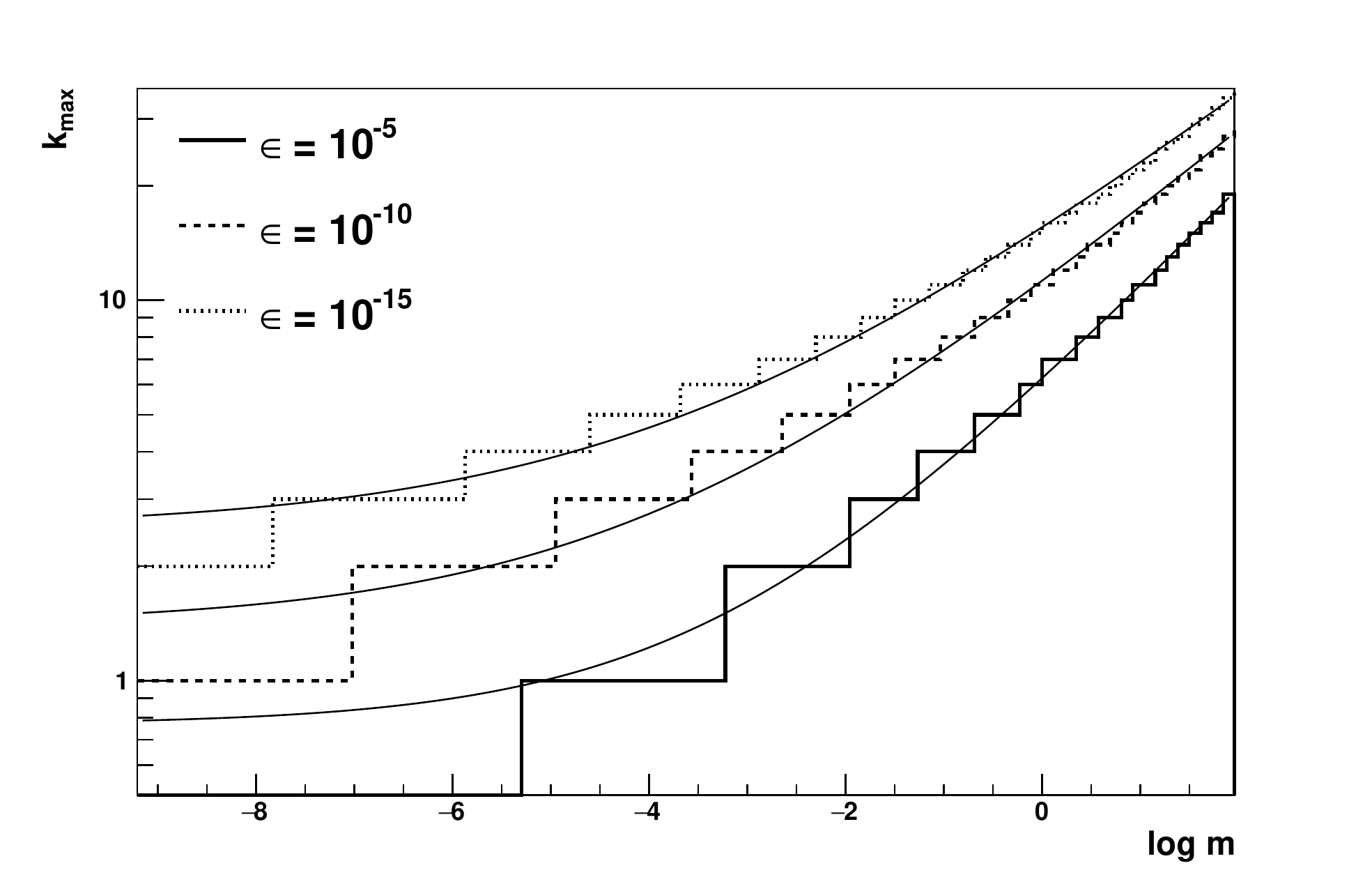}
  \caption{Fit of $k_\mathrm{max}$ as a function of $\log m$ for three choices of $\epsilon$ ($10^{-5}, 10^{-10}, 10^{-15}$).}
  \label{fig:showPoissonIntervalupper}
\end{figure}

For $m<5$, we set $k_\mathrm{min}(m,\epsilon) = 0.$ In order to determine $k_\mathrm{max}(m,\epsilon)$, we first compute $k_\mathrm{max}$ as a function of $\log m$ for several values of $\epsilon$, and fit these curves with the function $a+bm^{c}$, as shown in Figure~\ref{fig:showPoissonIntervalupper}.

\begin{figure}[ht]
  \centering
  \includegraphics[width=9.5cm]{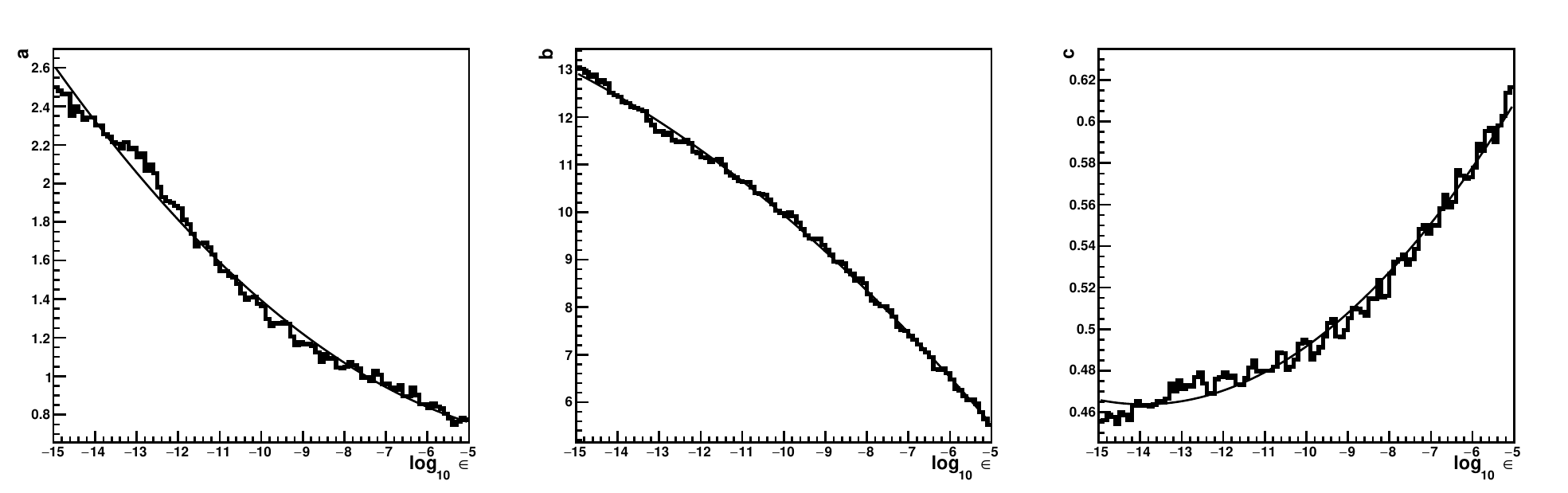}
  \caption{Variations with $\log_{10}\epsilon$ of the parameters of the function $a+bm^{c}$ used to fit the dependence of $k_\mathrm{max}$ with $m$.}
  \label{fig:showPoissonIntervalupperparam}
\end{figure}

We then look at the variation of the parameters $a,b$ and $c$ as a function of $\log_{10}\epsilon$ between $-15$ and $-5$. We fit these curves with a simple quadratic function, as shown in Figure~\ref{fig:showPoissonIntervalupperparam}. We find the following parametrizations:
\begin{align*}
  a(\epsilon) & =  0.715029+0.049825 \log_{10}\epsilon +0.011768(\log_{10}\epsilon)^2 \\
  b(\epsilon) & =  -0.308206-1.309547 \log_{10}\epsilon -0.028455(\log_{10}\epsilon)^2 \\
  c(\epsilon) & =  0.817286+0.050841 \log_{10}\epsilon+0.001828(\log_{10}\epsilon)^2
\end{align*}

\begin{figure}[ht]
  \centering
  \includegraphics[width=9.5cm]{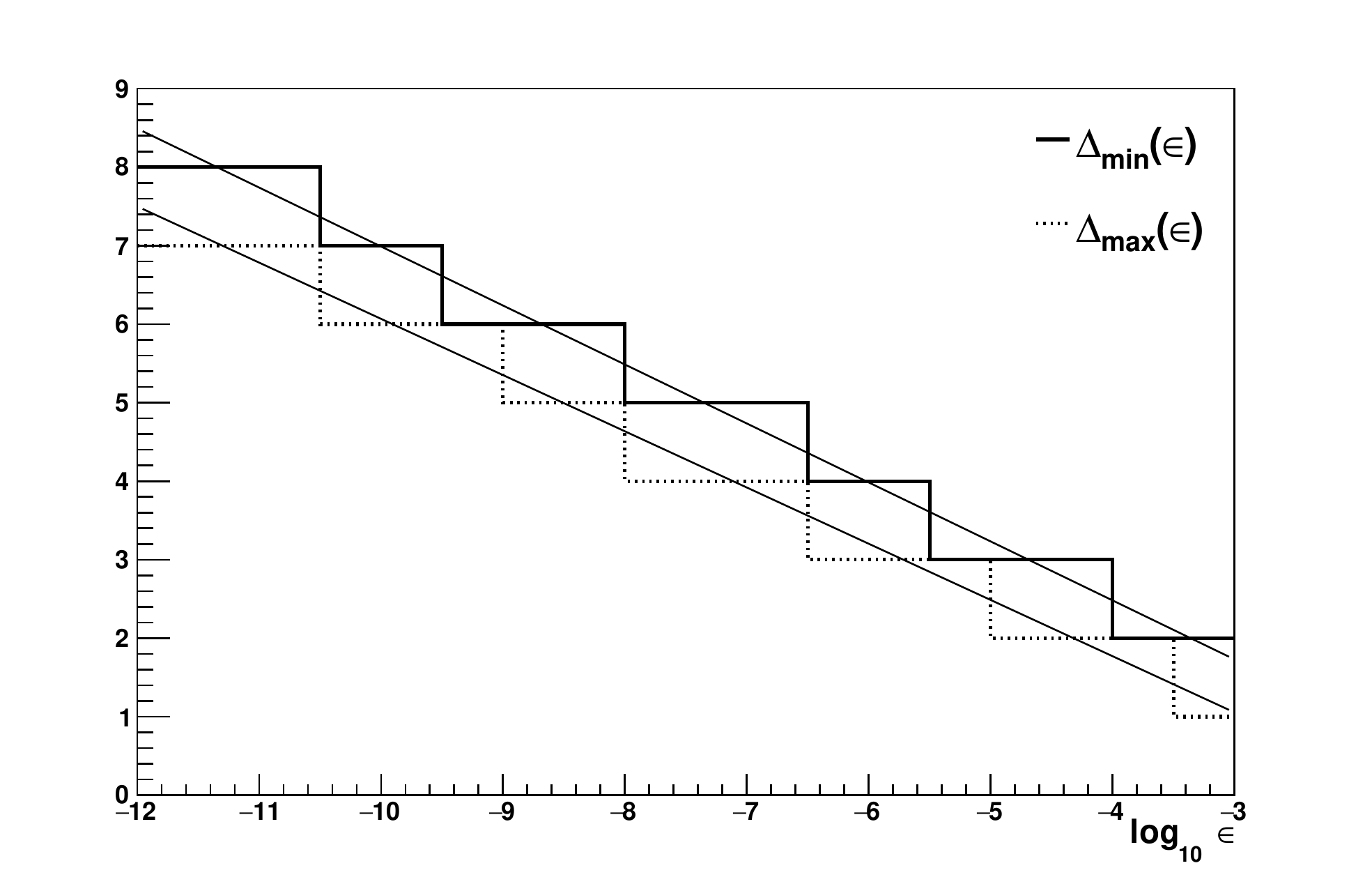}
  \caption{$\Delta_\mathrm{min}$ and $\Delta_\mathrm{max}$, the differences between $k_\mathrm{min}$ and $k_\mathrm{max}$ with respect to their Gaussian expectations, as a function of $\log_{10}\epsilon$.}
  \label{fig:showPoissonIntervalupper_lo_up}
\end{figure}

For $m \geq 5$, in order to find the narrowest $[k_\mathrm{min},k_\mathrm{max}]$ interval such that $\sum_{k} \mathcal{P}(k,m) \geq 1-\epsilon$, we sort the Poisson probabilities $\mathcal{P}(k,m)$ in decreasing order, find the smallest subset for which $\sum_{k} \mathcal{P}(k,m) \geq 1-\epsilon$ and then find the lowest and highest $k$ among this subset. Rather than trying to directly parameterize their variations with $m$ and $\epsilon$, we compare them to the Gaussian expectations $m \pm \sigma(\epsilon) \sqrt{m}$, with $\sigma(\epsilon) = \sqrt{2}\mathrm{erf}^{-1}(1-\epsilon)$. Because of the positive skewness of the Poisson distribution, both $k_\mathrm{min}$ and $k_\mathrm{max}$ are greater than their Gaussian expectations. We find that, for a given $\epsilon$, these differences are almost constant ($\pm$ 1 unit) for $m<200$. So we define:
\begin{align*}
  \Delta_\mathrm{min}(\epsilon) & =  \mathrm{min}_{5<m<200} \left( k_\mathrm{min} - (m - \sigma(\epsilon) \sqrt{m}) \right) \\
  \Delta_\mathrm{max}(\epsilon) & =  \mathrm{max}_{5<m<200} \left( k_\mathrm{max} - (m + \sigma(\epsilon) \sqrt{m}) \right) 
\end{align*}
Figure~\ref{fig:showPoissonIntervalupper_lo_up} shows that the variation of $\Delta_\mathrm{min}$ and $\Delta_\mathrm{max}$ with $\log_{10}\epsilon$ can be approximated with a linear parameterization. We use:
\begin{align*}
  k_\mathrm{min}(m,\epsilon)  & = \mathrm{max} \left( 0, m - \sigma(\epsilon) \sqrt{m} -0.523564-0.75129 \log_{10}\epsilon \right) \\
  k_\mathrm{max}(m,\epsilon)  & = m + \sigma(\epsilon) \sqrt{m} -1.09374-0.716202 \log_{10}\epsilon
\end{align*}

\section{Testing the weighted log-likelihood with simulations}  \label{app:systwithweights}

When computing the PS, we use Equation~\ref{eq:wloglike} to take into account the likelihood weights in the log-likelihood function. These weights are associated to a certain level of systematic uncertainty. As explained in Section~\ref{sec:wLL}, the absolute meaning of the systematic uncertainty is ensured for the spectral bins with Gaussian statistics by using the $\chi^2$ approximation to compute the corresponding $L$ pdf.

In order to study the case of the spectral bins with Poisson statistics, we perform simulations in which the number of observed counts in bin $k$ is drawn according to a Poisson distribution of mean $\mu_k$, which is itself drawn, for each realization, according to a Gaussian distribution of mean $m_k$ and standard deviation $\sigma m_k$, where $\sigma$ is the systematic uncertainty level. This simple simulation is not realistic because it ignores the likely correlations between spectral bins but it corresponds to a situation in which the correction brought by the $\chi^2$ approximation works exactly for Gaussian statistics. Therefore this simulation allows us to compare how well the weighted log-likelihood correction of Equation~\ref{eq:wloglike} performs for Poisson statistics relatively to the corrected $\chi^2$ approximation.

A simple analytical calculation shows that the mean and variance of the number of observed counts in each spectral bin are $m_k$ and $m_k+(\sigma m_k)^2$, respectively. It means that systematic uncertainties lower than 10\% have a negligible impact on spectral bins with $m_k\leq 1$.

\begin{figure}[ht]
  \centering
  \includegraphics[width=9.5cm]{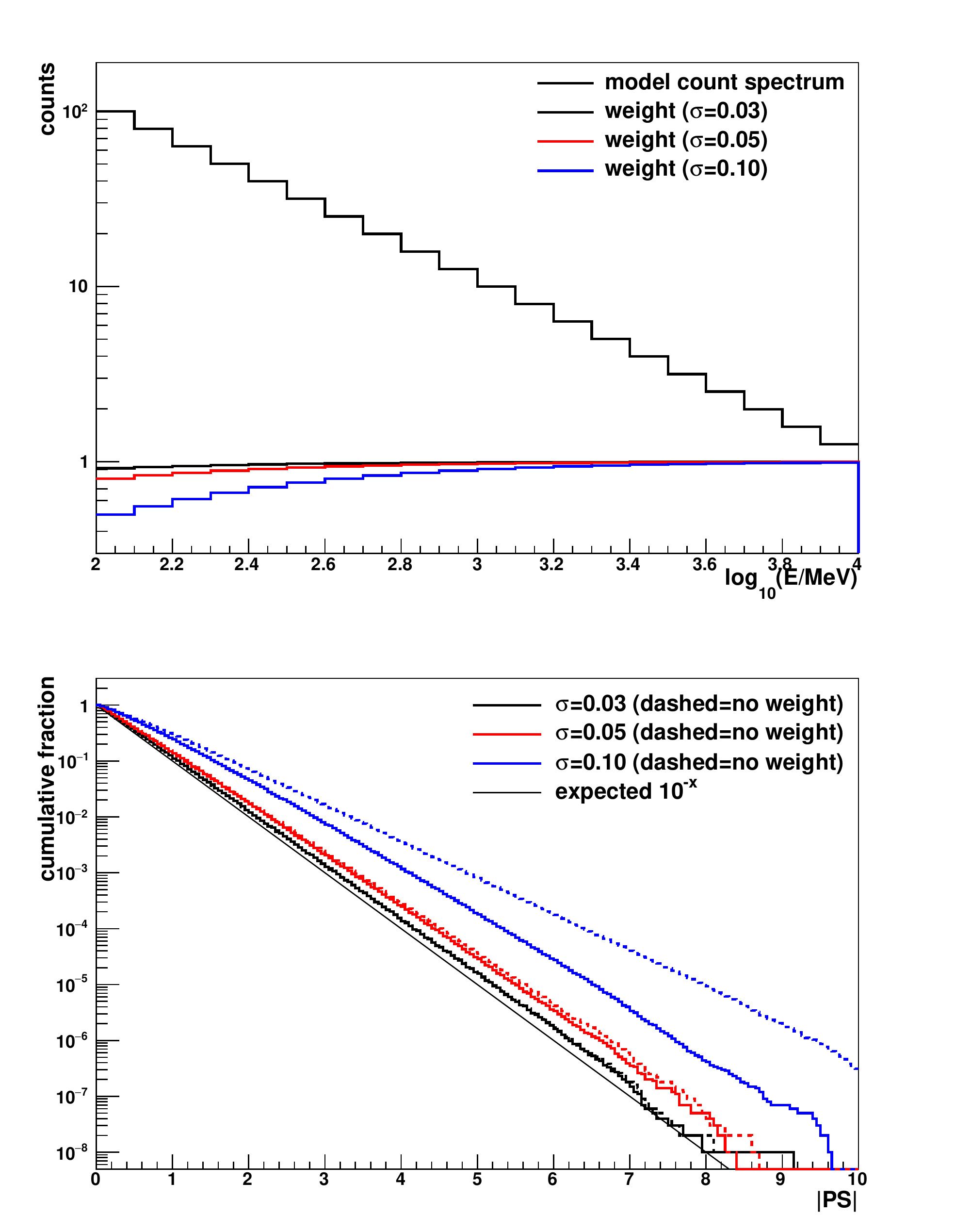}
  \caption{Simulations to test the use of likelihood weights to handle systematic uncertainty: count spectrum and average weights (top) and PS CCDF as a function of $|\mathrm{PS}|$ (bottom). The three levels of systematic uncertainty are 3, 5 and 10\%. The solid and dashed histograms correspond to the weighted and unweighted PS, respectively. The expected $10^{-x}$ distribution is also shown (thin solid line).}
  \label{fig:poissonwithsyst}
\end{figure}

We consider a 20~bins $E^{-2}$ spectrum such that the number of counts decreases from 100 to 1 and perform simulations for three different values of $\sigma$: 3, 5 and 10\%. We set the weights to $w_k = 1/(1+m_k\sigma^2)$. The count spectrum and the weights are shown in the top panel of Figure~\ref{fig:poissonwithsyst} and the resulting PS CCDF are shown in the bottom panel, as well as the ones obtained when no weight correction is applied.

If the weight correction were perfect, the PS CCDF should follow the $10^{-x}$ behavior. In the $\sigma=10$\% case, the weight correction has a significant effect but the error on PS is about 25\%. For $\sigma=3$\%, corresponding to the current LAT case, the weight correction has no effect but the error on PS is within 3\%. In the $\sigma=5$\% intermediate case, the weight correction starts to play a role but the resulting error on PS is about 10\%.

It appears from this study that, although the weighted version of the log-likelihood does not fully encompass the effect of systematic uncertainty, it allows the PS estimator to take it partly into account in the case of systematic uncertainty larger than 5\%.

\section{PS map production steps}  \label{app:PSproductionsteps}

In this Apppendix we recap the several steps that we go through to produce the PS map of a given RoI from the input data and model 3D count maps as well as the likelihood weight maps.

The data and model 3D count maps are produced with the {\tt Fermitools}\footnote{\url{https://fermi.gsfc.nasa.gov/ssc/data/analysis/scitools/references.html}} {\tt gtbin} and {\tt gtmodel}, respectively. The spatial part of the maps is a $12\degr \times 12\degr$ map with a pixel size of $0.1\degr$, while the energy part ranges from 100~MeV to 1~TeV with a $\log_{10}{E}$ bin size of 0.1. The likelihood weights are produced with {\tt gtwtsmap} from a 3D data count map covering a larger spatial region ($22.6\degr \times 22.6\degr$ with a pixel size of $0.1\degr$) in order to ensure a good estimation of the weights within the $12\degr \times 12\degr$ inner region that is used in the PS computation.

The three steps of the PS production, whose flowchart is shown in Figure~\ref{fig:diagramPSmap}, are the following:

\begin{itemize}
\item {\bf Spatial integration:} this step produces the 3D integrated count spectra. It is performed independently for each energy bin $k$, whose lower bound is noted $E_k$. For each $(i,j)$ pixel, we sum all the counts in the pixels within a distance $d(E_k)$ from pixel $(i,j)$, with $d(E_k) = 4\degr (E_k/100~\mathrm{MeV})^{-0.9} \oplus 0.1\degr$. For the log-likelihood weight 3D map, we compute the average weight over the pixels within a distance $d(E_k)$ from pixel $(i,j)$, as explained in Section~\ref{sec:wLL}.
\item {\bf Energy rebinning:} this step produces 3D integrated count spectra with a $\log_{10}{E}$ bin size of 0.3 from the 3D maps produced in the previous step. For each $(i,j)$ pixel, the energy bins of the data and model counts are summed by group of three, while for the log-likelihood weights we compute the average weight over the three merged bins (with the same prescription as explained in Section~\ref{sec:wLL}). The energy part of the output 3D maps ranges from 100~MeV to $10^{5.9}$~MeV with a $\log_{10}{E}$ bin size of 0.3.
\item {\bf PS computation:} this step produces the 2D PS map. For each $(i,j)$ pixel, we use the data and model integrated count spectra as well as the log-likelihood weights from the 3D maps produced in the previous step. The model integrated count spectra and the weights are used to compute the $L$ pdf (starting with the pdf corresponding to all the spectral bins with Gaussian statistics and then performing the iterative computation described in Appendix~\ref{app:iterativecomputation} for the remaining bins), which allows us to compute the p-value corresponding to the likelihood obtained with the data integrated counts. The absolute value of the pixel PS is $-\log_{10}(\text{p-value})$ and its sign is given by Equation~\ref{eq:signwPS}.

\begin{figure}[ht]
  \centering
  \includegraphics[width=9.5cm]{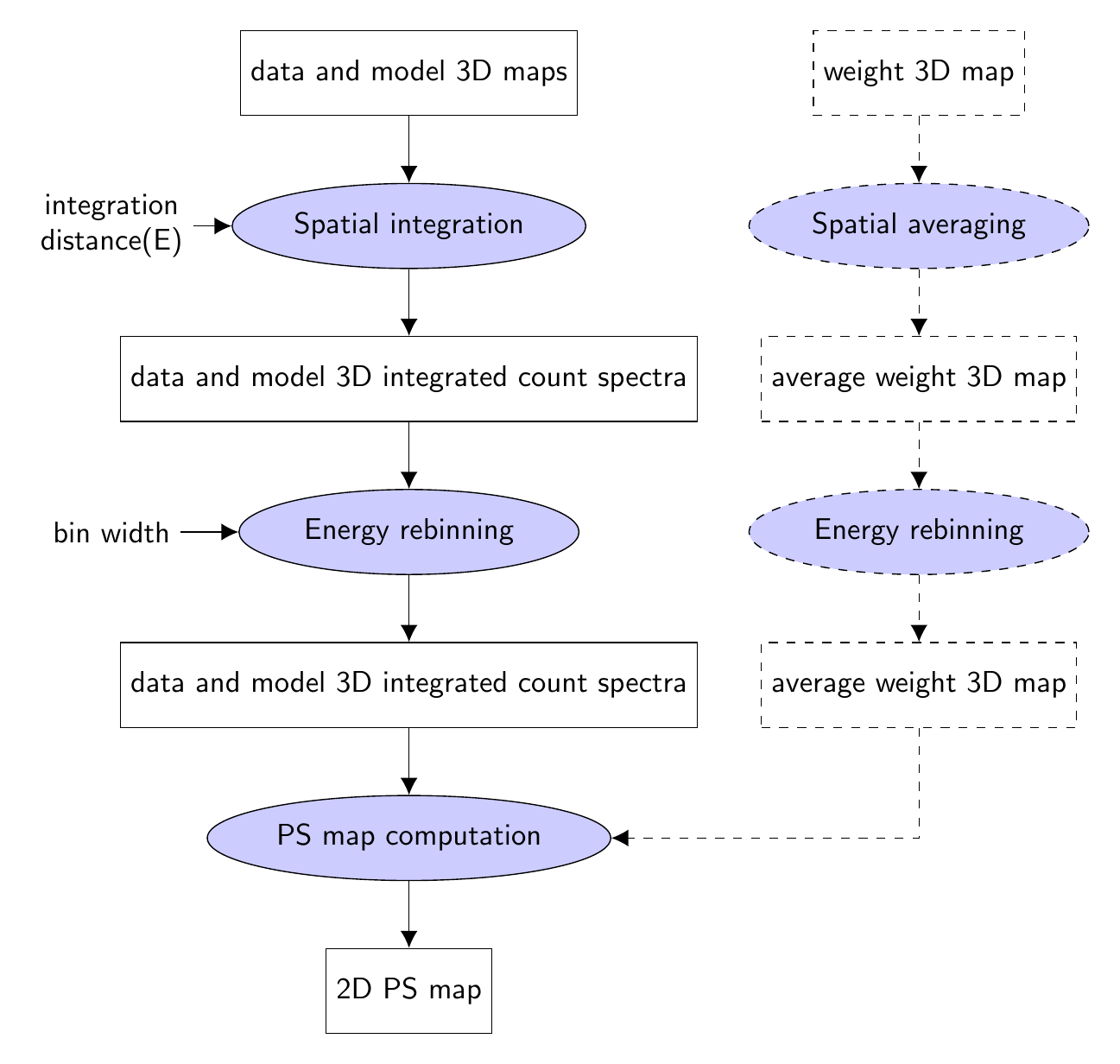}
  \caption{Flowchart of the PS map production.}
  \label{fig:diagramPSmap}
\end{figure}

\end{itemize}

\end{appendix}


\begin{thebibliography}{}

\bibitem[Abdo et al.(2009)]{LS5039_2009} Abdo, A. A., Ackermann, M., Ajello, M., et al. 2009, ApJ, 706, L5
\bibitem[Abdollahi et al.(2020)]{4FGL} Abdollahi, S., Acero, F., Ackermann, M., et al. 2020, ApJS, 247, 33
\bibitem[Ackermann et al.(2012)]{IRFs} Ackermann, M., Ajello, M., Albert, A., et al. 2012, ApJS, 203, 4
\bibitem[Ackermann et al.(2013)]{PSFDETERMINATION} Ackermann, M., Ajello, M., Allafort, A., et al. 2013, ApJ, 765, 1
\bibitem[Aydi et al.(2020)]{NOVA_paper} Aydi, E., Sokolovsky, K. V., Chomiuk, L., et al. 2020, Nature Astronomy, 4, 776–780
\bibitem[Atwood et al.(2009)]{latinstrument} Atwood, W. B., Abdo, A. A., Ackermann, M., et al. 2009, ApJ, 697, 1071
\bibitem[Atwood et al.(2013)]{pass8} Atwood, W., Albert, A., Baldini, L., et al. 2013, eConf C121028, 8, in Proc. 4th Fermi Symposium, Monterey
\bibitem[Ballet et al.(2020)]{4FGL-DR2} Ballet, J., Burnett, T. H., Digel, S. W., et al. 2020, arXiv:2005.11208
\bibitem[Bruel et al.(2018)]{P8R3} Bruel, P., Burnett, T. H., Digel, S. W., et al. 2018, presented at the 8th Fermi symposium, arXiv:1810.11394
\bibitem[Chang et al.(2016)]{LS5039_2016} Chang, Z., Zhang, S., Ji, L., et al. 2016, MNRAS, 463, 495
\bibitem[G\'orski et al.(2005)]{healpix} G\'orski, K. M., Hivon, E., Banday, A. J., et al. 2005, ApJ, 622, 759
\bibitem[Hadasch et al.(2012)]{LS5039_2012} Hadasch, D., Torres, D. F., Tanaka, T., et al. 2012, ApJ, 749, 54
\bibitem[Hu et al.(2002)]{loglikeweights} Hu, F., \& Zidek, J. V. 2002, Canad. J. Statist., 30, 347
\bibitem[Jean et al.(2018)]{NOVA_ATel_fermi} Jean, P., Cheung, C. C., Ojha, R., et al., The Astronomer’s Telegram, 11546
\bibitem[Mattox et al.(1996)]{mattox} Mattox, J. R., Bertsch, D. L., Chiang, J., et al. 1996, ApJ, 461, 396
\bibitem[Stanek et al.(2018)]{NOVA_ATel_optical} Stanek, K. Z., Holoie, T. W.-S., Kochanek, C. S., et al., The Astronomer’s Telegram, 11454
\bibitem[Wood et al.(2017)]{fermipy} Wood, M., Caputo, R., Charles, E, et al. 35th International Cosmic Ray Conference. 10-20 July, 2017. Bexco, Busan, Korea, Proceedings of Science, Vol. 301.
\bibitem[Yoneda et al.(2021)]{LS5039_2021} Yoneda, H., Khangulyan, D., Enoto, T. et al. 2021, submitted to ApJ
  
\end{thebibliography}
\end{document}